\pdfoutput=1
\pdfoutput=1
\documentclass[acmtog]{acmart}
\acmSubmissionID{1656}
\AtBeginDocument{%
  \providecommand\BibTeX{{%
    \normalfont B\kern-0.5em{\scshape i\kern-0.25em b}\kern-0.8em\TeX}}}

\newcommand{\highlight}[1]{\textcolor{blue}{#1}} 
\definecolor{MyRed}{rgb}{0.65,0.07,0.09}

\definecolor{MyGreen}{rgb}{0.18,0.55,0.09}

\definecolor{MyRed}{rgb}{1.0,0.0,0.0}
\newcommand{\revise}[1]{#1}
\usepackage{tcolorbox}
\usepackage{enumitem}
\usepackage{wrapfig}
\usepackage{diagbox}
\usepackage{verbatim}
\usepackage{amssymb}
\usepackage{latexsym}
\usepackage{lineno}
\usepackage{xspace}
\usepackage{pdfpages}
\usepackage{booktabs}  
\usepackage{color}
\graphicspath{{figures/}}
\usepackage{amsmath,bm}
\usepackage{psfrag}
\usepackage{mathtools}

\theoremstyle{definition}

\usepackage{multirow, tabularx}
\newcolumntype{Y}{>{\centering\arraybackslash}X}
\usepackage{capt-of}%
\usepackage{array, boldline, makecell, booktabs}

\newcommand{\Mod}[1]{\ (\mathrm{mod}\ #1)}
\usepackage{array}
\usepackage{pifont}
\usepackage{psfrag}
\usepackage{makecell}
\usepackage{algorithm}
\usepackage{threeparttable,booktabs}
\usepackage{graphicx}
\usepackage{svg}
\usepackage{subcaption}
\usepackage{flushend}
\usepackage{stfloats} 
\usepackage{appendix}
\usepackage[noend]{algpseudocode}

\usepackage{xcolor}
\usepackage{graphicx}

\newif\ifverbose
\verbosetrue

\ifverbose

\newcommand{\vb}[1]{\textcolor{red}{#1}}
\else

\newcommand{\vb}[1]{}
\fi

\usepackage{url}
\usepackage{xcolor}
\definecolor{newcolor}{rgb}{.8,.349,.1}

\copyrightyear{2025}
\acmYear{2025}
\setcopyright{cc}
\setcctype{by}
\acmConference[SA Conference Papers '25]{SIGGRAPH Asia 2025 Conference Papers}{December 15--18, 2025}{Hong Kong, Hong Kong}
\acmBooktitle{SIGGRAPH Asia 2025 Conference Papers (SA Conference Papers '25), December 15--18, 2025, Hong Kong, Hong Kong}\acmDOI{10.1145/3757377.3763903}
\acmISBN{979-8-4007-2137-3/2025/12}

\begin{document}
\author{Zhiqi Li}
\authornotemark[1]
\email{zli3167@gatech.edu}
\affiliation{
\institution{Georgia Institute of Technology}
\country{USA}
}

\author{Jinjin He}
\authornote{Joint first author}
\email{jhe433@gatech.edu}
\affiliation{
\institution{Georgia Institute of Technology}
\country{USA}
}

\author{Barnab\'as B\"orcs\"ok}
\email{borcsok@gatech.edu}
\affiliation{
\institution{Georgia Institute of Technology}
\country{USA}
}

\author{Taiyuan Zhang}
\email{taiyuan.zhang.gr@dartmouth.edu}
\affiliation{
\institution{Dartmouth College}
\country{USA}
}

\author{Duowen Chen}
\email{dchen322@gatech.edu}
\affiliation{
\institution{Georgia Institute of Technology}
\country{USA}
}

\author{Tao Du}
\email{taodu.eecs@gmail.com}
\affiliation{
\institution{Independent Researcher}
\country{~}
}

\author{Ming C. Lin}
\email{lin@umd.edu}
\affiliation{
\institution{University of Maryland}
\country{USA}
}

\author{Greg Turk}
\email{turk@cc.gatech.edu}
\affiliation{
\institution{Georgia Institute of Technology}
\country{USA}
}

\author{Bo Zhu}
\email{bo.zhu@gatech.edu}
\affiliation{
\institution{Georgia Institute of Technology}
\country{USA}
}

\title{An Adjoint Method for Differentiable Fluid Simulation on Flow Maps}

\begin{abstract}
This paper presents a novel adjoint solver for differentiable fluid simulation based on bidirectional flow maps. Our key observation is that the forward fluid solver and its corresponding backward, adjoint solver share the same flow map \revise{as} the forward simulation. In the forward pass, this map transports fluid impulse variables from the initial frame to the current frame to simulate vortical dynamics. In the backward pass, the same map propagates adjoint variables from the current frame back to the initial frame to compute gradients. This shared long-range map allows the accuracy of gradient computation to benefit directly from improvements in flow map construction. Building on this insight, we introduce a novel adjoint solver that solves the adjoint equations directly on the flow map, enabling long-range and accurate differentiation of incompressible flows without differentiating intermediate numerical steps or storing intermediate variables, as required in conventional adjoint methods. To further improve efficiency, we propose a long-short time-sparse flow map representation for evolving adjoint variables. Our approach has low memory usage, requiring only 6.53GB of data at a resolution of $192^3$ while preserving high accuracy in tracking vorticity, enabling new differentiable simulation tasks that require precise identification, prediction, and control of vortex dynamics.
\end{abstract}



\begin{teaserfigure}
\centering%
\includegraphics[width=1.0\textwidth]{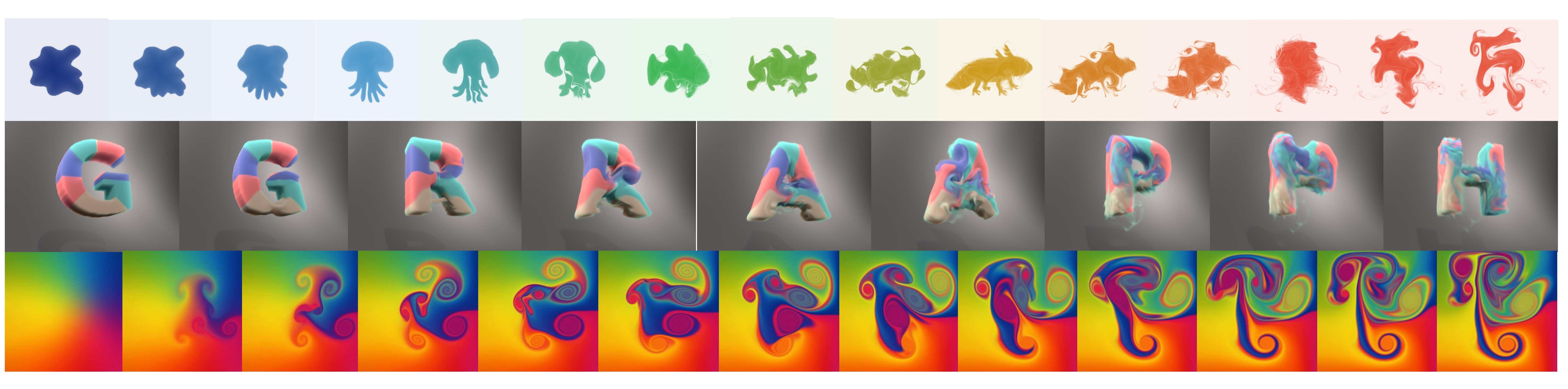}
\caption{{Demonstration of our differentiable fluid simulation on flow maps using our adjoint solver:} \textit{Top:} A sequence of 2D fluid shape optimizations demonstrating smooth morphing between target silhouettes. \textit{Middle:} 3D fluid control with multiple keyframes to guide a 3D letter morphing from "G" to "R" to "A" to "P" to "H". \textit{Bottom:} A vortex dynamics inference task that predicts future flow evolution from a sequence of observed past images.}
 \label{fig:teaser}
\end{teaserfigure}

\maketitle

\section{Introduction}


Accurately differentiating a dynamic fluid system, particularly computing the derivatives of fluid variables over a long time horizon, remains a fundamental challenge in both computer graphics and computational physics. The primary difficulty arises from the intrinsic flow nature of fluids: unlike solid systems with more constrained configurations, fluid systems evolve freely over space and time under physical laws, giving rise to a high-dimensional, continuously deforming state space that significantly complicates the backward differentiation process. As the simulation progresses over longer time periods (either forward or backward), numerical errors accumulate at each timestep, further degrading the accuracy of gradient computation and making the long-horizon derivative estimation increasingly unreliable.

Two mainstream approaches have been developed for differentiating fluid systems governed by the Navier–Stokes equations in both computer graphics and computational physics. One class of methods directly differentiates the discrete numerical scheme used in the forward simulation. A representative example is the pioneering work by \citet{mcnamara2004fluid}, which employed a classical advection-projection scheme \cite{stam1999stable} in the forward process and computes gradients by sequentially differentiating the advection and projection steps, where an adjoint system is solved to account for the projection. This method has been highly successful within the graphics community and has inspired a substantial body of follow-up work (e.g., see \citet{li2024neuralfluid,holl2024phiflow,takahashi2021differentiable}).
Since this approach directly targets the discrete formulation used in the forward solver, it ensures that the computed gradients are fully consistent with the actual simulation steps, which is critical for using gradient information to guide optimization processes (e.g., in control, animation, or design problems). The other class of methods derives the adjoint system analytically at the level of the continuous governing equations, followed by discretization of the resulting adjoint PDEs. Such approaches have been widely adopted in computational fluid dynamics for solving inverse problems (e.g., see \citet{stuck2012adjoint,galecki2022adjoint}). However, to ensure that the discrete adjoint solution accurately corresponds to the derivative of the discrete forward simulation, these methods often rely on high-order discretization, particularly in their advection schemes and spatial operators, which limits their practicality in visual computing scenarios where computational efficiency and scalability are critical.

We propose a new adjoint solver that improves both the accuracy and efficiency of existing methods for differentiable incompressible flow simulation. Our approach is built upon the concept of long-range bidirectional flow maps, which have recently emerged as an effective modeling framework for simulating a wide range of fluid systems and their multiphysics couplings dominated by vortical dynamics (e.g., see \cite{deng2023fluid,zhou2024eulerian,li2024particle,chen2024solid} for examples). The core idea of the flow-map method is to construct a mapping between the initial time and the current time that accurately transports physical quantities between corresponding spatial locations. The term "bidirectional" refers to the capability of transporting quantities both forward and backward in time, with the forward and backward maps forming a consistent and temporally symmetric pair. A key observation underlying our work is that this bidirectional flow map, originally introduced to enhance the accuracy of forward simulation, can be naturally repurposed to support the backward adjoint process. Sharing the same flow map across both forward and backward processes enables temporally symmetric transport of fluid quantities and their adjoints over extended time intervals, which is a proven strength of flow-map-based formulations. Leveraging this accuracy, our method eliminates the need to differentiate individual numerical steps, thereby avoiding the high memory and computational overhead associated with discrete differentiation, and instead enables direct solution of the continuous adjoint PDEs with improved scalability and precision.

Motivated by this idea, we developed a novel adjoint solver grounded in flow map theory to enable long-range, accurate differentiation of incompressible flow systems. Our system comprises three key components: (1) a forward incompressible fluid solver based on bidirectional flow maps discretized over a sequence of grid-aligned frames; (2) a backward adjoint solver that solves the adjoint equations using the same flow maps \revise{as in the forward process}; and (3) an acceleration strategy based on a long-short time-sparse flow map representation to reduce computational cost without sacrificing accuracy. The forward and backward solvers not only share the same grid-based bidirectional flow maps but also apply the same numerical scheme to evolve fluid quantities and their adjoints, respectively, along opposite time directions. 


\section{Related Work}

\paragraph{Differentiable Fluid Simulation}

Differentiable fluid simulation in computer graphics typically computes gradients by differentiating the discretized forward simulation. The pioneering works by \revise{\citet{treuille2003keyframe} and \citet{mcnamara2004fluid}} differentiates the discretized advection-projection fluid simulation method \cite{stam1999stable} and has inspired a line of subsequent works \revise{\cite{holl2020phiflow, holl2024phiflow, li2024neuralfluid, takahashi2021differentiable, pan2017efficient}}. Previous works have also studied differentiating smoothed-particle hydrodynamics (SPH) \cite{li2023difffr}, reduced-mode fluids \cite{chen2024fluid}, and the lattice-boltzmann method (LBM) \cite{ataei2024xlb}. These techniques have enabled progress in fluid control and optimization, but they can suffer from limited accuracy or high memory cost when applied to long-horizon simulations. In contrast, our work does not differentiate a discretized fluid simulator but directly discretize the continuous adjoint PDEs of the Navier-Stokes equations, enabling more accurate numerical solutions to the adjoint equations.

\paragraph{Adjoint Methods} The adjoint method is a standard mathematical tool for computing gradients in PDE-constrained optimization. Studies in computational fluid dynamics (CFD) typically apply it to derive the adjoint Navier-Stokes equations \revise{\cite{jameson1988aerodynamic, giles2003algorithm, giles2000introduction, stuck2012adjoint}}, enabling sensitivity analysis through backward-time evolution. \revise{Beyond fluids, adjoint formulations underpin a broad class of differentiable physics frameworks, where gradients of physical systems are leveraged for inverse design, control, and optimization. Applications span elastic materials \cite{geilinger2020add, du2021diffpd, hu2019chainqueen, qiao2021differentiable}, cloth simulation \cite{li2022diffcloth, qiao2020scalable}, contact and collision \cite{huang2024differentiable, huang2023optimized}, magnetic shells \cite{chen2022simulation}, and topology optimization \cite{liu2018narrow, sigmund200199, zhu2017two, feng2023cellular}. These differentiable physics systems enable diverse applications including robot design \cite{ma2021diffaqua, gjoka2024soft}, surface optimization \cite{mehta2022level, montes2023differentiable, he2024multi}, parameter identification \cite{hahn2019real2sim, li2023pac, ma2022risp}, microstructure discovery \cite{sigmund200199, huang2023optimized}, and policy learning \cite{qiao2021efficient, huang2021plasticinelab, xian2023fluidlab,li2018learning}, typically using gradient-based optimizers like Adam \cite{kingma2014adam} or LBFGS \cite{nocedal1999numerical}.}

\paragraph{Flow Map Methods} Flow map methods trace their origins to the method of characteristic mapping (MCM) by \citet{wiggert1976numerical}, later developed in computer graphics by \citet{tessendorf2011characteristic} and \citet{qu2019efficient}. 
Recent progress on representing a bidirectional map includes neural network-based storage compression \cite{deng2023fluid}, buffer-free Eulerian representations \cite{li2025edge}, and the particle flow map method \cite{li2024particle, zhou2024eulerian, li2025clebsch, chen2025fluid, wang2025cirrus}, which have further enhanced accuracy. Gauge-based fluid formulations \cite{buttke1992lagrangian, cortez1996impulse} have been explored with various applications \cite{feng2022impulse, nabizadeh2022covector,li2024lagrangian}. Despite their accuracy advantages, flow map methods suffer from high computational complexity, with traditional \revise{Eulerian flow map (EFM) method \cite{deng2023fluid} }requiring $O(n^2)$ flow map evolution costs, recently addressed by time-sparse approaches \cite{Sun_2025}.

\begin{figure}[t]
    \centering
    \includegraphics[width=1.0\linewidth]{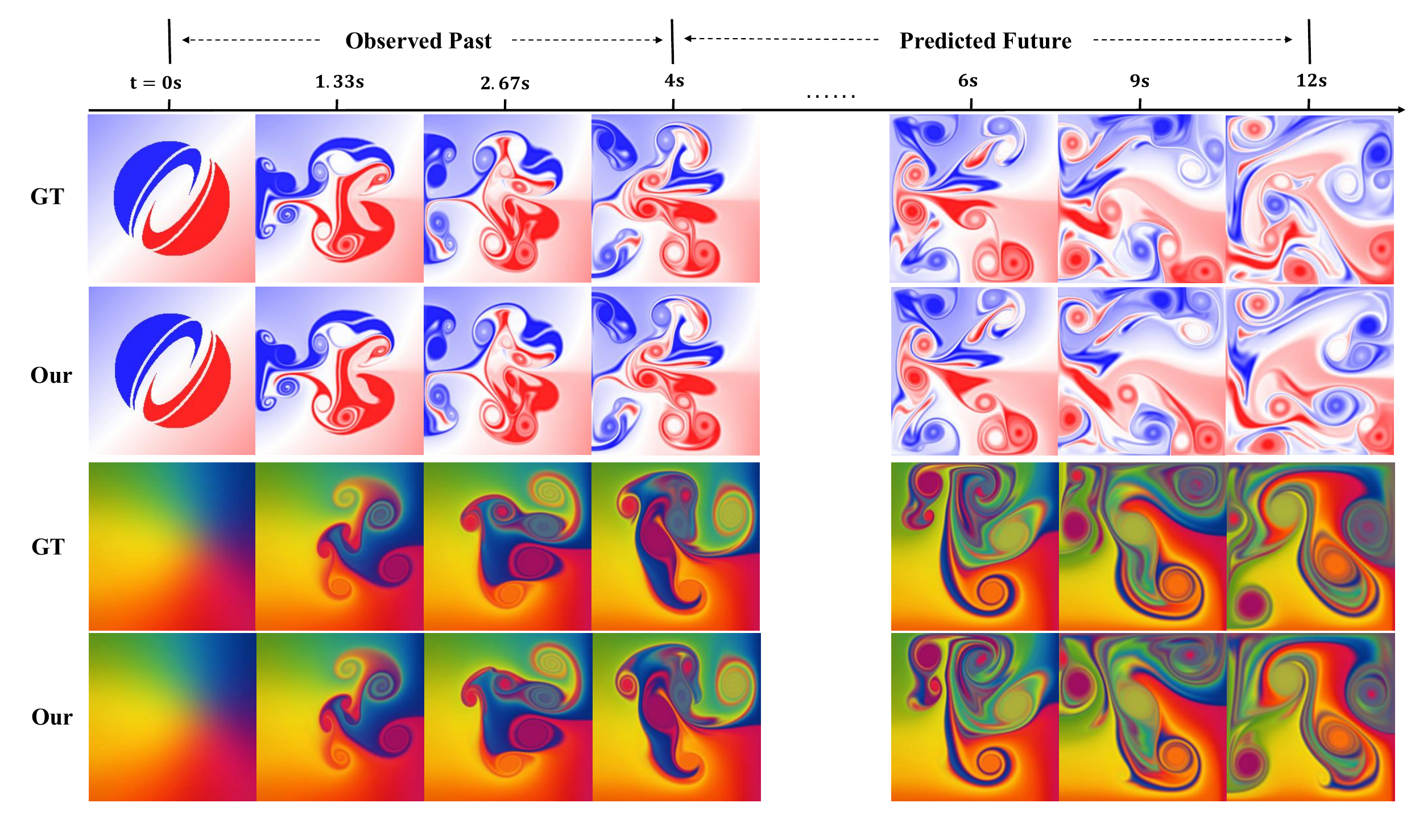}
    \caption{\textbf{Vortex dynamics inference from velocity-field videos.} Training on the first 4 seconds infers 8 random vortices, maintaining accuracy over extended 12-second predictions.}
    \label{fig:2DLogo_and_Gradient}
\end{figure}

\section{Physical Model}
\subsection{Differentiable Fluid}
\paragraph{Fluid Equations}
We focus on the incompressible Navier–Stokes equations and the advection of a passive field for fluid simulation:
\begin{equation}\label{eq:NS_equation}
    \begin{aligned}
\left( \frac{\partial}{\partial t} + (\mathbf{u} \cdot \nabla) \right) \mathbf{u} &= -\frac{1}{\rho} \nabla p + \nu \Delta \mathbf{u} + \mathbf{f}, \\
\nabla \cdot \mathbf{u} &= 0,
\end{aligned}
\end{equation}
\begin{equation}\label{eq:NS_equation2}
    \begin{aligned}
\left( \frac{\partial}{\partial t} + (\mathbf{u} \cdot \nabla) \right) \xi &= 0\revise{,}
\end{aligned}
\end{equation}
where $p$, $\mathbf{f}$, $\nu = \frac{\mu}{\rho}$ denote the pressure field, external force, and the kinematic viscosity, respectively. The scalar field $\xi$ represents a passive quantity field advected by the fluid, such as smoke density or color. Given the velocity field $\mathbf{u}_{s'}$ and passive field $\xi_{s'}$ at arbitrary start time $s'\ge s $, \autoref{eq:NS_equation} determines their exact evolution for any $t \ge s'$ with well-defined boundary conditions, denoted as $(\mathbf{u}_t, \xi_t) = \mathbf{F}_{s'\to t}(\mathbf{u}_{s'}, \xi_{s'})$.  The fluid simulation method approximates this process numerically as $(\hat{\mathbf{u}}_t, \hat{\xi}_t) = \hat{\mathbf{F}}_{s'\to t}(\mathbf{u}_{s'}, \xi_{s'})$, where $\hat{\cdot}$ denotes numerical approximation.

\begin{figure}[t]
    \centering
    \includegraphics[width=1.0\linewidth]{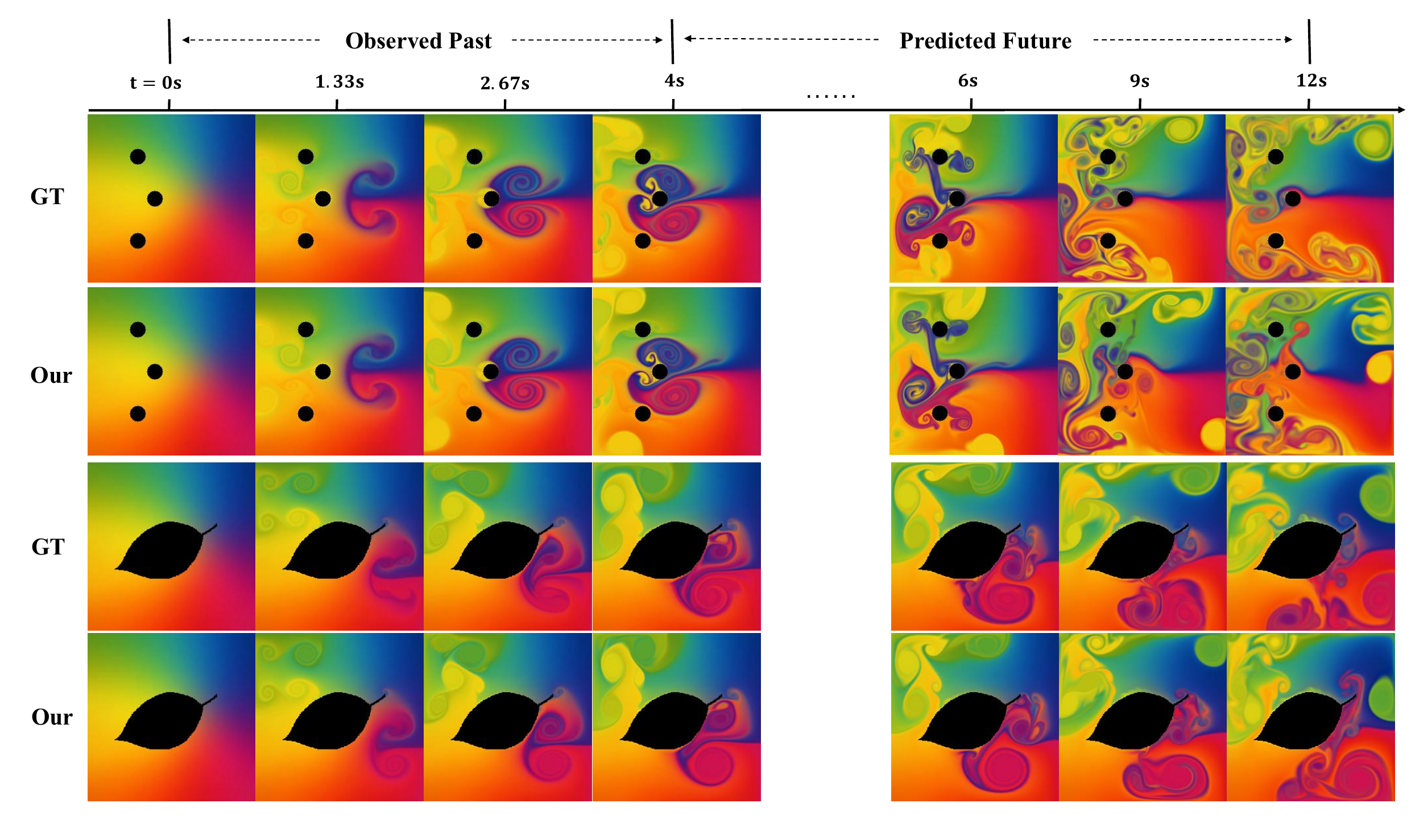}
    \caption{\textbf{Vortex dynamics with obstacle interference.} Method successfully infers vortices with vortex-obstacle interactions, enabling accurate long-term flow prediction around geometric constraints.}
    \label{fig:2Dobstacal}
\end{figure}

\paragraph{Adjoint Equations}
In fluid-related optimization problems, we aim to minimize an objective functional 
\begin{equation}\label{eq:object_function_def}
    \begin{aligned}
        L(\mathbf{u},\xi) = \int_{s}^{r} \int_{\mathbb{U}_t} J(\mathbf{u},\xi, t) d\mathbf{x} dt,
    \end{aligned}
\end{equation}
where \revise{the objective functional integrand} $J$ is a time-dependent functional of the velocity field $\mathbf{u}$ and the passive scalar $\xi$, for example, the terminal velocity loss $J(\mathbf{u},\xi,t) = \delta(t - r) \| \mathbf{u} - \mathbf{u}_{\text{target}} \|_2^2$, where $\mathbf{u}_{\text{target}}$ and $\delta$  denote the target velocity field and the Dirac delta function, respectively.  When applying common optimization methods such as gradient descent to minimize $L$, it is necessary to compute the gradient information $\mathbf{u}_t^* = \frac{\partial L}{\partial \mathbf{u}_t}$ and $\xi_t^* = \frac{\partial L}{\partial \xi_t}$, which are referred to as the adjoints of $\mathbf{u}_t$ and $\xi_t$, respectively.  \cite{stuck2012adjoint, galecki2022adjoint} shows that $\mathbf{u}_t^*$ and $\xi_t^*$ follow the equations:
\begin{equation}\label{eq:adjoint_NS_equation}
    \begin{aligned}
\left( \frac{\partial}{\partial t} + (\mathbf{u} \cdot \nabla) \right) \mathbf{u}^* &=\nabla \mathbf{u}^\top \mathbf{u}^*+ \xi^*\nabla \xi -\frac{1}{\rho} \nabla p^* + \nu \Delta \mathbf{u}^* - \frac{\partial J}{\partial \mathbf{u}},  \\
\nabla \cdot \mathbf{u}^* &= 0, \\
\left( \frac{\partial}{\partial t} + (\mathbf{u} \cdot \nabla) \right) \xi^* &= -\frac{\partial J}{\partial \xi},
\end{aligned}
\end{equation}
where $p^*$ is the adjoint pressure. Here we assume that the external force $\mathbf{f}$ is independent of $\mathbf{u}$ and $\xi$, and more general cases can be derived using the adjoint method.  Given the adjoint velocity field $\mathbf{u}^*_{r'}$ and passive field $\xi^*_{r'}$ at time $r'\le r$, \revise{\autoref{eq:adjoint_NS_equation}} determine their exact solution for any $t \le r'$ with boundary conditions, denoted as $(\mathbf{u}^*_t, \xi^*_t) = \mathbf{B}_{r'\to t}(\mathbf{u}^*_{r'}, \xi^*_{r'})$.  Differentiable fluid simulation aims at approximating this process numerically as $(\hat{\mathbf{u}}^*_t, \hat{\xi}^*_t) = \hat{\mathbf{B}}_{r'\to t}(\mathbf{u}^*_{r'}, \xi^*_{r'})$.

\begin{figure}[htbp]
  \centering
  \begin{subfigure}[b]{0.2\textwidth}
    \includegraphics[width=\textwidth]{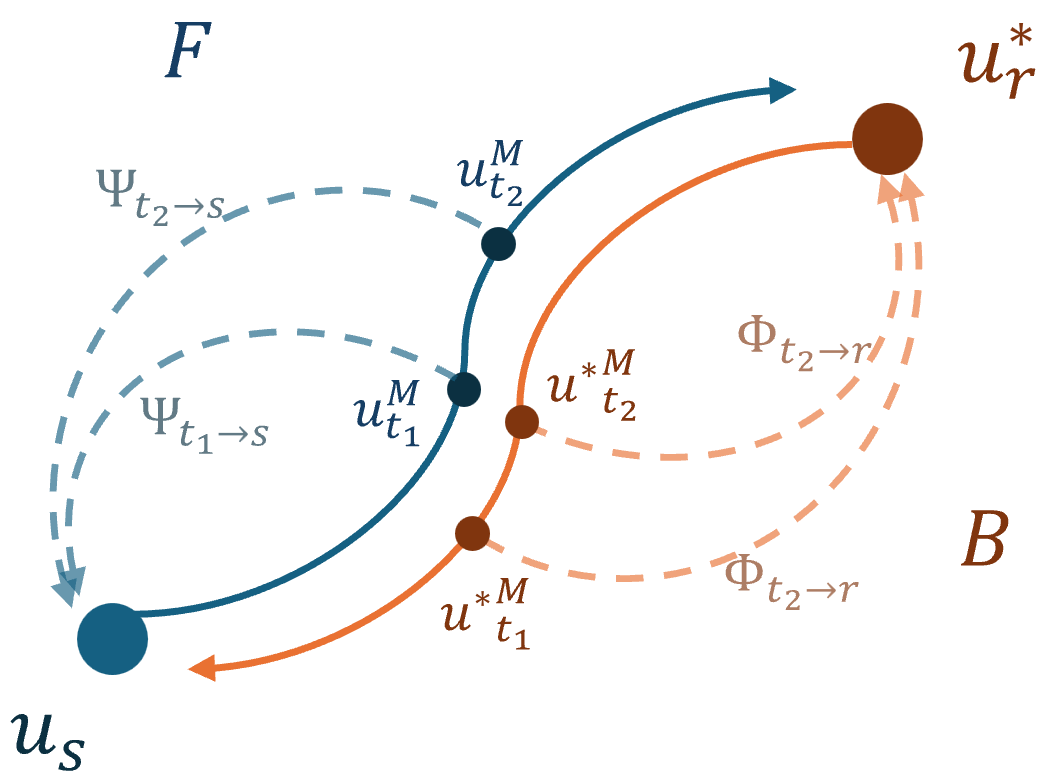}
    \caption{}
  \end{subfigure}
  \hfill
  \begin{subfigure}[b]{0.2\textwidth}
    \includegraphics[width=\textwidth]{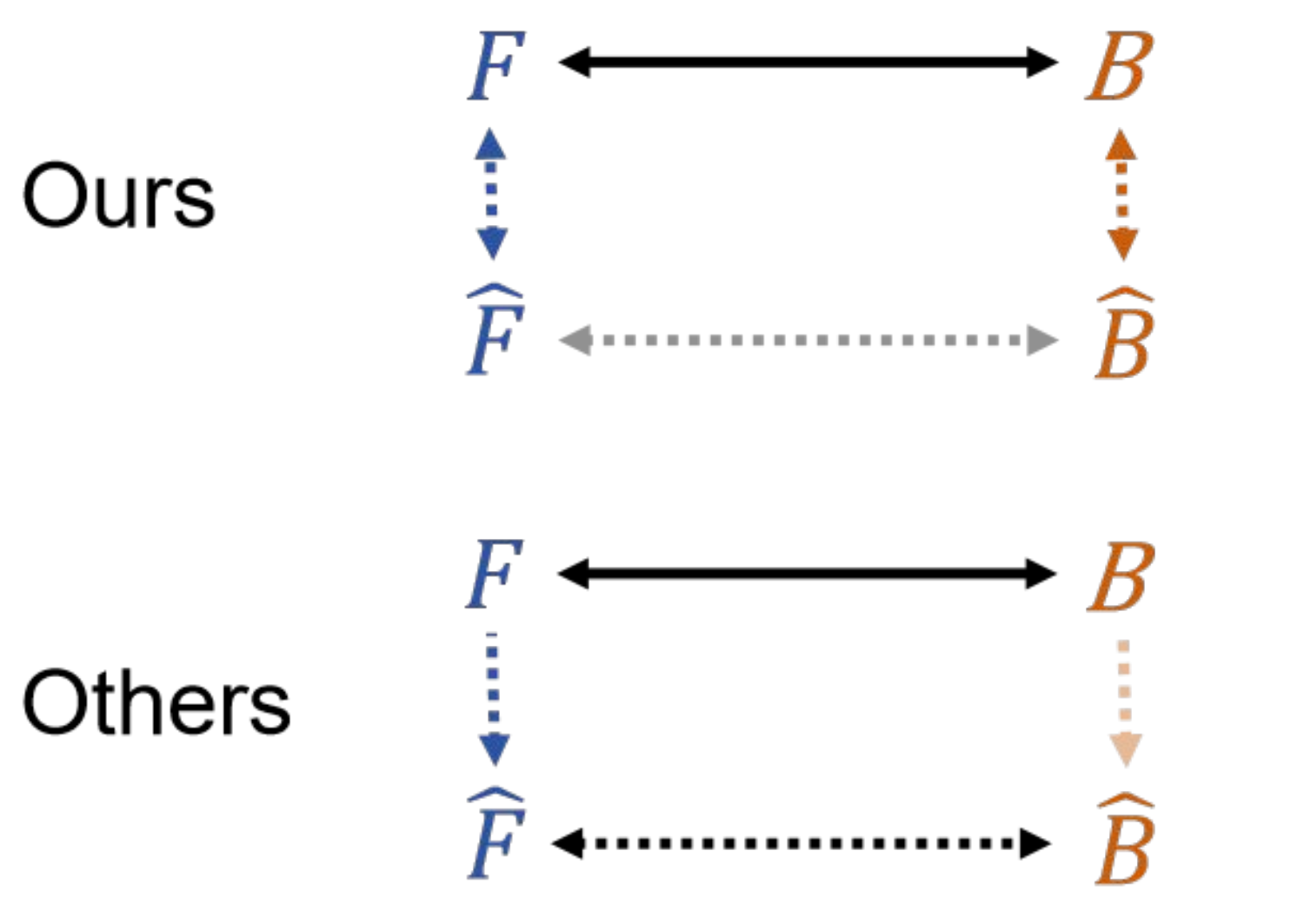}
    \caption{}
  \end{subfigure}
  \vspace*{-0.5em}
  \caption{\textbf{Method overview.} In (a), we illustrate the symmetry between the forward and backward passes. The forward pass maps $\revise{\mathbf{u}}$ using the backward flow map $\revise{\mathbf{\Psi}}$, while the backward pass maps $\revise{\mathbf{u}}^*$ using the forward flow map $\revise{\mathbf{\Phi}}$. Both $\revise{\mathbf{\Psi}}$ and $\revise{\mathbf{\Phi}}$ are opposite to the flow direction and require repeated long-range integration for accuracy, leading to the original EFM’s $O(m^2)$ time complexity, where $m$ is the flow map length.  In (b), we compare our method with other differentiable approaches. Due to lower accuracy, \revise{existing methods indirectly approximate $\mathbf{B} \to \hat{\mathbf{B}}$ (semi-transparent one-way arrows) through approximating $\mathbf{F} \to \hat{\mathbf{F}}$ (dashed one-way arrows) and directly differentiating $\hat{\mathbf{F}}$ (dashed double arrows).  Although $\hat{\mathbf{B}}$ is consistent with $\hat{\mathbf{F}}$, $\mathbf{B} \to \hat{\mathbf{B}}$ is inaccurate. }  In contrast, our method leverages the strict correspondence between $\revise{\mathbf{F}}$ and $\revise{\mathbf{B}}$.  We only need to construct accurate approximations \revise{$\mathbf{F}\to\mathbf{\hat{F}}$ and $\mathbf{B}\to\mathbf{\hat{B}}$} respectively (dashed double arrows), and the consistency between $\revise{\mathbf{\hat{F}}}$ and $\revise{\mathbf{\hat{B}}}$ then naturally follows through transitivity (semi-transparent dashed double arrows), \revise{enabled by the higher accuracy of flow maps.} }
  \label{fig:multi_caption}
  \vspace*{-1em}
\end{figure}

\subsection{Method Overview}

\revise{We address fluid-related optimization problems using the flow map method, aiming to optimize parameters $\theta$ so that the resulting fluid states $\mathbf{u}^\theta_t(\mathbf{x})$ and $\xi^\theta(\cdot)$ minimize the objective functional $L(\mathbf{u}^\theta, \xi^\theta)$ under specific control scenarios and subject to constraints of fluid dynamics (\autoref{eq:NS_equation}). The process iteratively performs forward simulation $\hat{\mathbf{F}}_{s\to r}$, evaluates the functional $L(\mathbf{u}^\theta, \xi^\theta)$, computes the backward adjoint process $\hat{\mathbf{B}}_{r\to s}$ to obtain adjoints $\mathbf{u}_t^{*\theta}$ and $\xi_t^{*\theta}$, and updates the control parameters $\theta$ using these adjoints.}

\revise{To obtain the scheme for $\hat{\mathbf{B}}_{r\to s}$, unlike previous differentiable fluid solvers that differentiate the discretized forward process $\hat{\mathbf{F}}_{s\to t}$, our approach directly computes the continuous backward process $\mathbf{B}_{r\to t}$ via flow maps, yielding a principled adjoint formulation and establishing a symmetric forward–backward framework that uses flow maps for the consistent evolution of states and adjoints. A high-level comparison and overview of this computation are shown in Fig.~\ref{fig:multi_caption}. We then introduce the flow map method (\autoref{sec:flowmap_introduction}), describe its use for forward (\autoref{sec:flowmap_forward}) and adjoint computation (\autoref{sec:flowmap_backward}), present a novel acceleration strategy (\autoref{sec:Time_Sparse_EFM}), and finally assemble the complete numerical scheme (\autoref{sec:numerical}), which we subsequently employ to solve fluid optimization problems (\autoref{sec:examples}).}
\begin{figure}[htbp]
  \centering
  \hspace{-0.5cm}
  \begin{subfigure}[b]{0.22\textwidth}
    \includegraphics[width=1.08\linewidth]{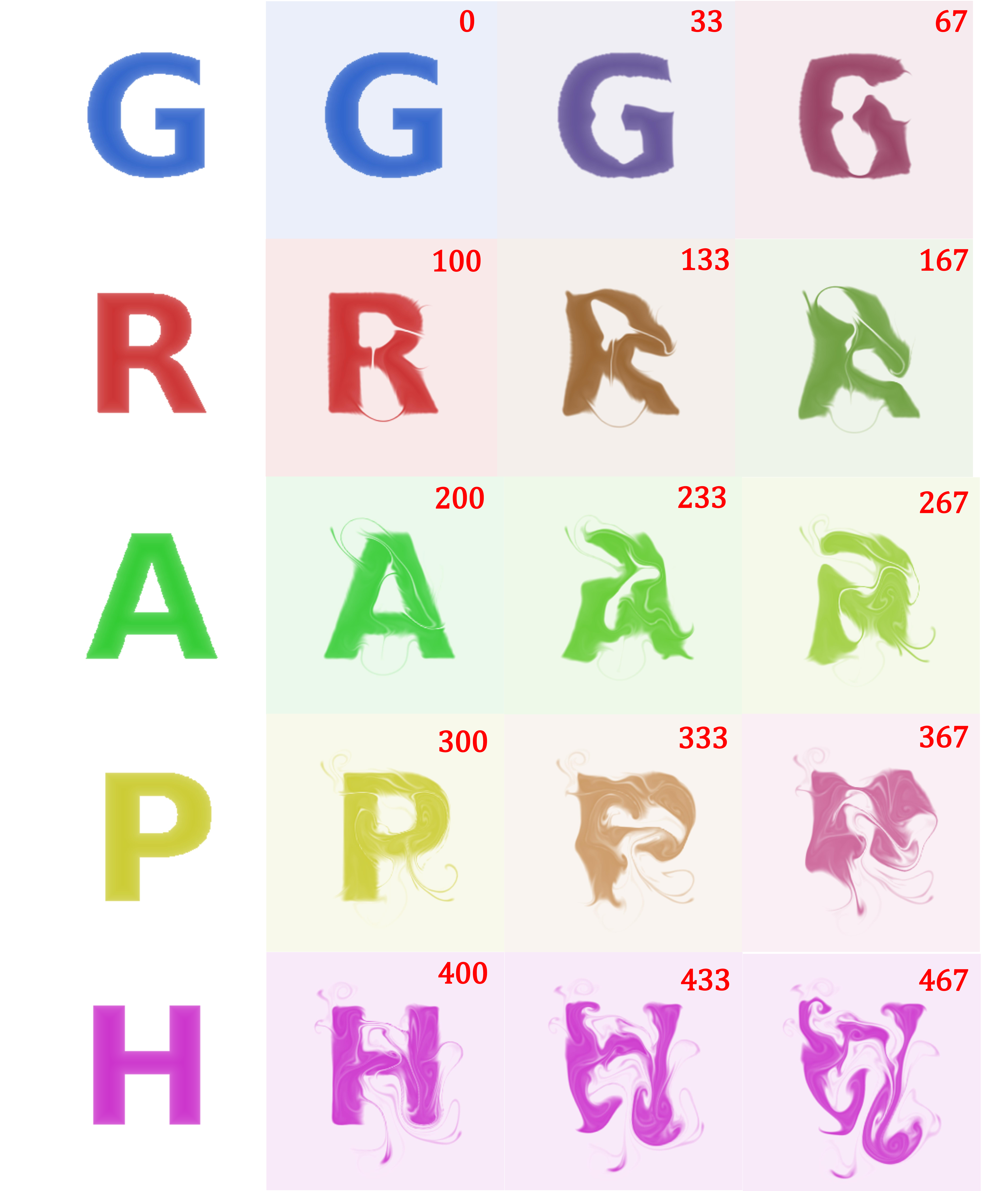}
    \caption{Letter Morphing}\label{fig:2DGRAPH}
  \end{subfigure}
  \hspace{0.25cm}
  \begin{subfigure}[b]{0.23\textwidth}
    \includegraphics[width=1.0\linewidth]{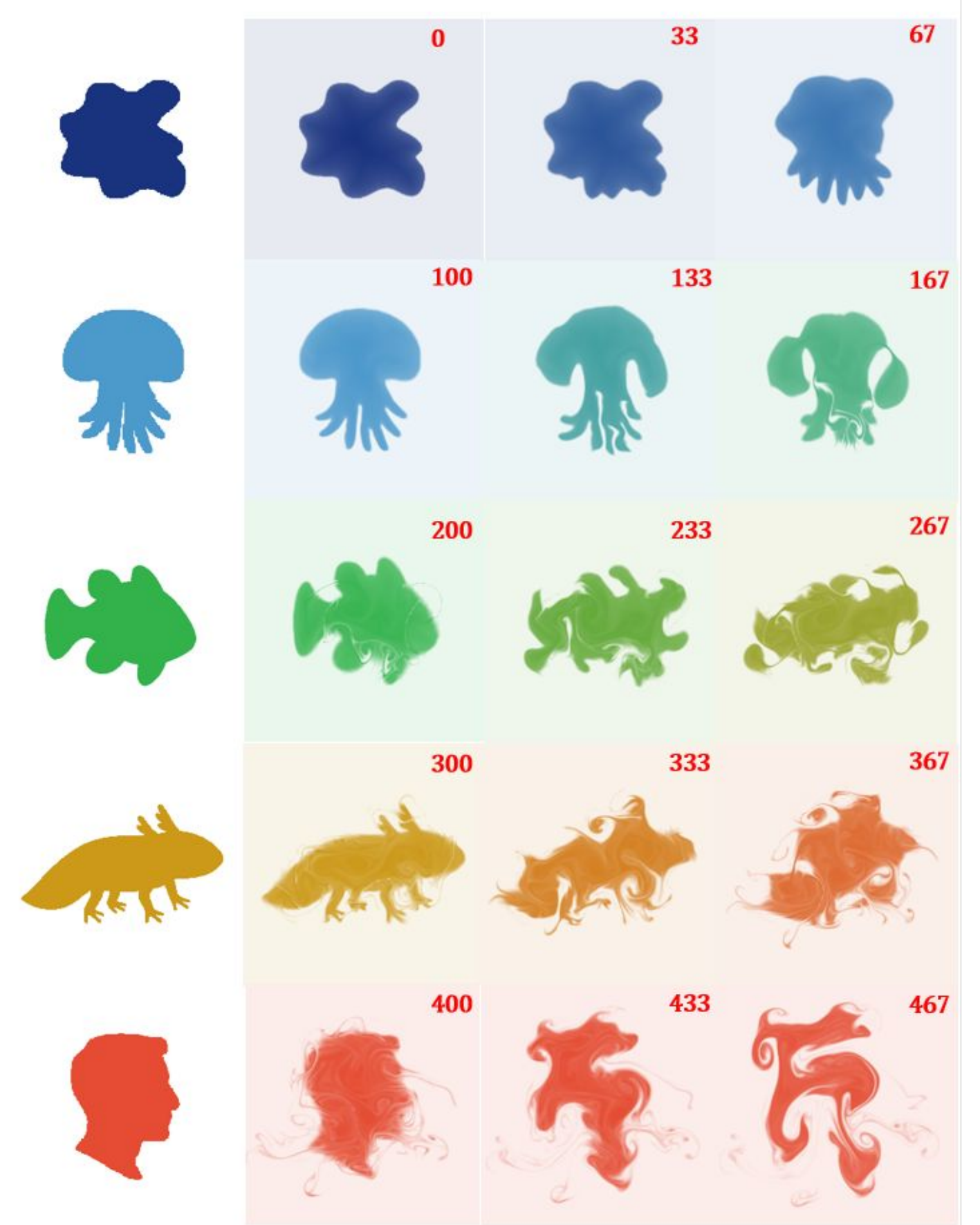}
    \hspace{-0.5cm}
    \caption{Life-Form Evolution}\label{fig:2Dlife}
  \end{subfigure}
  \vspace*{-0.5em}
  \caption{\textbf{2D sequential optimizations.} A sequence of 2D morphing tasks, including letter morphing ('G'→'R'→'A'→'P'→'H') and life-form evolution, demonstrating smooth transitions between target silhouettes. Each row illustrates the progressive transformation between two consecutive keyframes, with target shapes shown on the left.}
  \vspace*{-1em}
\end{figure}
\subsection{Flow Map}\label{sec:flowmap_introduction}
In fluid simulation,  the flow map method \cite{deng2023fluid,zhou2024eulerian} enables accurate advection of physics \revise{quantities} by constructing a mapping between the initial domain $\mathbb{U}_{s}$ and the current domain $\mathbb{U}_t, r > t > s$ where $s$ and $r$ are the initial time and the final time respectively.  Consider a fluid moving with a velocity field $\revise{\mathbf{u}(\mathbf{x},t)}$, $\revise{\mathbf{x}} \in \mathbb{U}_t$. For any time $t_1<t_2$, the forward flow map $\mathbf{\Phi}_{t_1\to t_2}: \mathbb{U}_{t_1} \to \mathbb{U}_{t_2}$ and backward flow map $\mathbf{\Psi}_{t_2\to t_1}: \mathbb{U}_{t_2} \to \mathbb{U}_{t_1}$ are defined as functions satisfying $\mathbf{\Phi}_{t_1\to t_2}(\mathbf{x}_q({t_1})) = \mathbf{x}_q({t_2})$ and $\mathbf{\Psi}_{{t_2}\to {t_1}}(\mathbf{x}_q({t_2})) = \mathbf{x}_q({t_1})$ for any fluid particle $q$ moving with $\frac{d\mathbf{x}_q(t)}{dt} = \mathbf{u}(\mathbf{x}_q(t), t)$, where $\mathbf{x}_q(t)$ \revise{denotes} position of particle $q$ at time $t$.  The Jacobian matrices of the flow maps are denoted as $\mathcal{F}_{t_1\to t_2}(\mathbf{x}) = \frac{\partial \mathbf{\Phi}_{t_1\to \revise{t_2}}(\mathbf{x})}{\partial \mathbf{x}}, \mathbf{x}\in \mathbb{U}_{t_1}$ and $\mathcal{T}_{t_2\to t_1}(\revise{\mathbf{x}}) = \frac{\partial \mathbf{\Psi}_{t_2\to t_1}(\revise{\mathbf{x}})}{\partial \revise{\mathbf{x}}}, \revise{\mathbf{x}}\in \mathbb{U}_{t_2}$, respectively.  From start time $s'\ge s$ selected arbitrarily, flow maps and their Jacobians follow evolution equations:
\begin{equation}\label{eq:Phi_F_advection}
\begin{cases}
    \frac{\partial \mathbf{\Phi}_{s' \to t}(\mathbf{x})}{\partial t} = \mathbf{u}(\mathbf{\Phi}_{s' \to t}(\mathbf{x}),t), & \mathbf{\Phi}_{s' \to s'}(\mathbf{x}) = \mathbf{x},\\
    \frac{\partial \mathcal{F}_{s' \to t}(\mathbf{x})}{\partial t} = \nabla \mathbf{u}(\mathbf{\Phi}_{s' \to t}(\mathbf{x}),t) \mathcal{F}_{s' \to t}(\mathbf{x}),&\mathcal{F}_{s' \to s'}(\mathbf{x}) = \mathbf{I},
\end{cases}
\end{equation}
\begin{equation}\label{eq:Psi_T_advection}
\begin{cases}
    \frac{D \mathbf{\Psi}_{t \to s'}(\mathbf{x})}{D t} = 0, & \mathbf{\Psi}_{s' \to s'}(\mathbf{x}) = \mathbf{x},\\
    \frac{D \mathcal{T}_{t \to s'}(\mathbf{x})}{D t} = -\mathcal{T}_{t \to s'}(\mathbf{x}) \nabla \mathbf{u}(\mathbf{x},t) ,&\mathcal{T}_{s' \to s'}(\mathbf{x}) = \mathbf{I}.
\end{cases}
\end{equation}

\begin{figure}[t]
    \centering
    \includegraphics[width=1.0\linewidth]{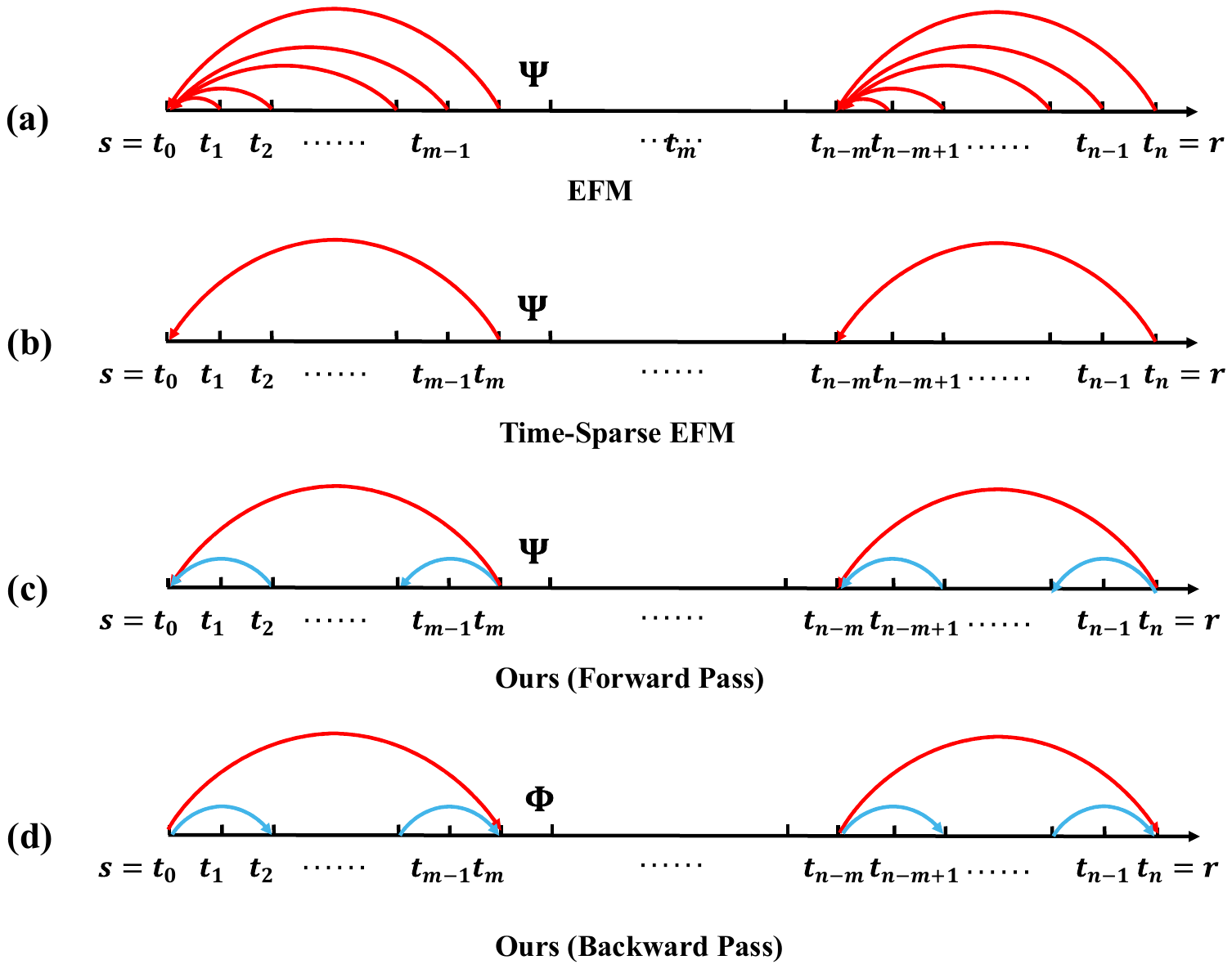}
    \caption{\textbf{Illustration of long-range flow map evolution in different methods.} Let $m$ be the reinitialization interval, typically $m = 15 \sim 60$.  (a) EFM computes $\mathbf{\Psi}_{t\to s'}$ at every time step by repeatedly evolving back to the previous reinit time. The number of steps crossed by each curve in the figure indicates the number of steps required for each flow map evolution, resulting in a total cost of $O(m^2)$.  (b) Time-Sparse EFM computes $\mathbf{\Psi}_{t\to s'}$ only at reinit steps, reducing the cost to $O(m)$.  (c)(d) Our improved Time-Sparse EFM introduces shorter intermediate flow maps between reinit steps to improve accuracy at intermediate times, while maintaining the overall cost at $O(m)$—specifically, doubling the number of flow map steps compared to (b).}
    \label{fig:different_mapping}
    \vspace*{-1em}
\end{figure}

In the flow map method, accurate calculation of advection depends on the accuracy of flow maps. While the forward flow map $\mathbf{\Phi}_{s'\to t}$ and its Jacobian $\revise{\mathcal{F}}_{s'\to t}$ can be integrated accurately on grids using high-order schemes, like \revise{the} Fourth-order Runge-Kutta method (RK4) for computing \autoref{eq:Phi_F_advection}, the semi-Lagrangian treatment of advection terms $\frac{D}{Dt}$ in \autoref{eq:Psi_T_advection} introduces dissipation and accumulates \revise{errors}, making precise computation of $\mathbf{\Psi}$ and $\mathcal{T}$ challenging.  To address this issue, \cite{deng2023fluid} observes that at any given time $r'$, the backward flow map $\mathbf{\Psi}_{r' \to s'}$ and its Jacobian $\mathcal{T}_{r' \to s'}$ can be interpreted as the result of evolving $\mathbf{\Psi}_{r' \to t}$ and $\mathcal{T}_{r' \to t}$ backward in time from $r'$ to $s'$ with $\Delta t <0$, which follows the dynamics of the reverse-time fluid motion without advection terms with start time $r'$ (see Fig.~\ref{fig:different_mapping} for illustration):
\begin{equation}\label{eq:nfm_evolution2}
    \begin{cases}
    \displaystyle \frac{\partial \mathbf{\Psi}_{r' \to t}(\mathbf{x})}{\partial t} = \mathbf{u}(\mathbf{\Psi}_{r' \to t}(\mathbf{x}),t), \quad & \mathbf{\Psi}_{r' \to r'}(\mathbf{x}) = \mathbf{x}, \\
    \displaystyle \frac{\partial \mathcal{T}_{r' \to t}(\mathbf{x})}{\partial t} = \nabla \mathbf{u}(\mathbf{\Psi}_{r' \to t}(\mathbf{x}),t)\, \mathcal{T}_{r' \to t}(\mathbf{x}), \quad & \mathcal{T}_{r' \to r'}(\mathbf{x}) = \mathbf{I}.
    \end{cases}
\end{equation}
Since \autoref{eq:nfm_evolution2} excludes advection terms, high-order integration can accurately compute the backward flow map $\mathbf{\Psi}$ and its Jacobian $\mathcal{T}$.  This approach is known as the Eulerian Flow Map method (EFM), and based on the accurately computed flow maps from \autoref{eq:Phi_F_advection} and \autoref{eq:nfm_evolution2}, it \revise{achieves} state-of-the-art performance in preserving vortex structures.

\section{Differentiable Flow Maps}

To implement differentiable flow maps, we compute $\revise{\hat{\mathbf{B}}}$ by directly discretizing the backward process $\revise{\mathbf{B}}$, rather than differentiating $\revise{\hat{\mathbf{F}}}$ as in previous methods. \revise{Using flow maps}, we first solve the Navier-Stokes \autoref{eq:NS_equation} forward from the start time $s$ to the end time $r$, then solve the adjoint Navier-Stokes \autoref{eq:adjoint_NS_equation} backward from $r$ to $s$. These two processes are referred to as the forward and backward pass, which will be discussed below.

\subsection{Forward Pass}\label{sec:flowmap_forward}
By \cite{li2024particle}, \autoref{eq:NS_equation} can be accurately computed \revise{via flow maps}, using \revise{the integral form} of \autoref{eq:NS_equation}
\begin{equation}\label{eq:velocity_solution_after_integration}
    \begin{aligned}
        &\mathbf{u}(\mathbf{x},t) = \mathcal{T}^\top_{t\to s}(\mathbf{x}) \mathbf{u}(\mathbf{\Psi}_{t\to s}(\mathbf{x}),s) + \mathcal{T}^\top_{t\to s}(\mathbf{x}) \mathbf{\Gamma}_{s\to t}(\mathbf{\Psi}_{t\to s}(\mathbf{x})),\\ 
        &\mathbf{\Gamma}_{s\to t}(\mathbf{x})= \int_{s}^t \mathcal{F}^\top_{s\to \tau}(\mathbf{x})\left(-\frac{1}{\rho} \nabla p+\frac{1}{2}\nabla \|\mathbf{u}\|^2 + \mathbf{f}\right)(\mathbf{\Phi}_{s\to \tau}(\mathbf{x}),\tau) d\tau,\\
         &\mathbf{\xi}(\mathbf{x},t) = \mathbf{\xi}(\mathbf{\Psi}_{t\to s}(\mathbf{x}),s).
    \end{aligned}
\end{equation}
The detailed procedure is omitted here and provided in Appendix A of the supplementary material. We follow the same approach to accurately compute the evolution of the adjoint.

\subsection{Backward Pass}\label{sec:flowmap_backward} 

Since the adjoint velocity field $\mathbf{u}^*_t$ also satisfies the incompressibility condition $\nabla \cdot \mathbf{u}^*_t = 0$, $\autoref{eq:adjoint_NS_equation}$ can similarly be solved using flow maps. Notably, both the forward \autoref{eq:NS_equation} and the adjoint \autoref{eq:adjoint_NS_equation} are driven by the same velocity field, and the flow of the backward pass can be viewed as the time reversal of the forward pass. As a result, the forward flow maps $\mathbf{\Phi}_{t_1\to t_2}$, $\mathcal{F}_{t_1\to t_2}$ and backward flow maps $\mathbf{\Psi}_{t_2\to t_1}$, $\mathcal{T}_{t_2\to t_1}$, $t_1<t_2$ used in the forward equations can serve as the backward and forward flow maps, respectively, in the backward pass of adjoint equations, allowing \revise{adjoint fields} $\mathbf{u}^*_t$ and $\boldsymbol{\xi}^*_t$ to be expressed as:
\begin{equation}\label{eq:adjoint_velocity_solution_after_integration}
    \begin{aligned}
        &\mathbf{u}^*(\mathbf{x},t) = \mathcal{F}^\top_{t\to r}(\mathbf{x}) \mathbf{u}^*(\mathbf{\Phi}_{t\to r}(\mathbf{x}),r) + \revise{\mathcal{F}}^\top_{t\to r}(\mathbf{x}) \mathbf{\Lambda}^u_{r\to t}(\mathbf{\Phi}_{t\to r}(\mathbf{x})),\\ 
        &\mathbf{\Lambda}^u_{r\to t}(\mathbf{x})= \int_{r}^t \mathcal{T}^\top_{r\to \tau}(\mathbf{x})\Big(2\nabla \mathbf{u}^\top \mathbf{u}^*+ \xi^*\nabla \xi -\frac{1}{\rho} \nabla p + \nu \Delta \mathbf{u}^* \\
        &\qquad\qquad\qquad\qquad\qquad- \frac{\partial J}{\partial \mathbf{u}}\Big)(\mathbf{\Psi}_{r\to \tau}(\mathbf{x}),\tau) d\tau,\\
        &\xi^*(\mathbf{x},t) =\xi^*(\mathbf{\Phi}_{t\to r}(\mathbf{x}),r) +  \mathbf{\Lambda}^\xi_{r\to t}(\mathbf{\Phi}_{t\to r}(\mathbf{x}))\revise{,}\\
        & \mathbf{\Lambda}^\xi_{r\to t}(\mathbf{x}) = - \int_r^t \frac{\partial J}{\partial \xi}(\mathbf{\Psi}_{r\to \tau}(\mathbf{x}),\tau) d\tau\revise{.}
    \end{aligned}
\end{equation}
Here, $\mathbf{u}^{*M}_{r \to t}(\mathbf{x}) = \mathcal{F}^\top_{t \to r}(\mathbf{x})\mathbf{u}^*(\mathbf{\Phi}_{t \to r}(\mathbf{x}), r)$ is referred to as the long-range mapped adjoint velocity, \revise{and $\mathbf{\Lambda}^u_{r \to t}$} denotes the path integrator of the adjoint velocity.  The long-range mapping allows $\mathbf{u^*}$ to avoid the error accumulation caused by advection.  For $\mathbf{\xi^*}(\mathbf{x},t)$, $\xi^{*M}_{r \to t}(\mathbf{x}) =\mathbf{\xi^*}(\mathbf{\Phi}_{t\to r}(\mathbf{x}),r)$ is the long-range mapped adjoint passive field and $\mathbf{\Lambda}_{r\to t}^\xi$ is its path integrator.

\begin{figure}[t]
    \centering
    \includegraphics[width=1.0\linewidth]{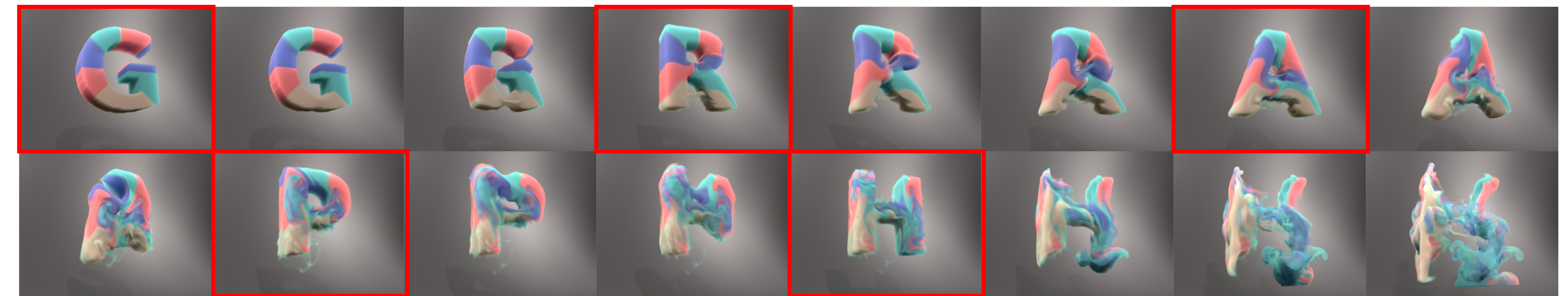}
    \caption{\textbf{3D sequential optimization.}  We perform 3D fluid control with multiple keyframes to guide a 3D shape morphing sequence from "G" to "R" to "A" to "P" to "H". The red boxes highlight keyframes where the fluid configuration successfully matches the target shapes.}
    \label{fig:3DGRAPH}
\end{figure}

\begin{figure}[t]
    \centering
    \includegraphics[width=1.0\linewidth]{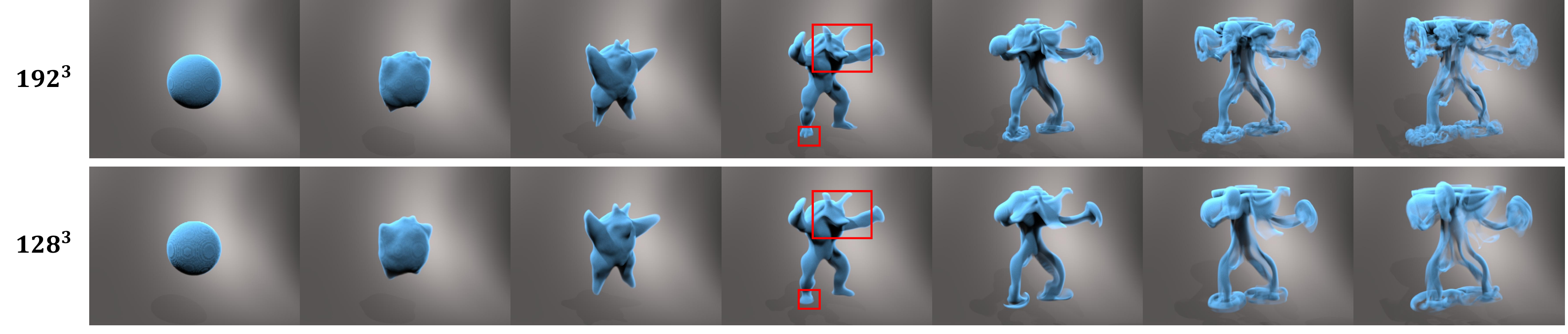}
    \caption{\textbf{Armadillo shape morphing at different resolutions.} We compare shape morphing results at two resolutions: $192^3$ (top) and $128^3$ (bottom). The initial sphere is progressively optimized to match the Armadillo shape. Higher resolution preserves finer geometric details, particularly in regions highlighted by red boxes.}
    \label{fig:3DArmadillo}
\end{figure}


For the adjoint calculation in the backward pass, we leverage the long-short term mapping conversion strategy introduced in \cite{chen2024solid, li2024particle} to formulate our strategy to calculate the adjoint integration \autoref{eq:adjoint_velocity_solution_after_integration}. We discuss the details as follows.

\paragraph{Mapping and Conversion} \revise{We} first compute the long-range mapped adjoint velocity $\mathbf{u}^{*M}_{r\to t}(\mathbf{x})$ using $\mathbf{u}^{*M}_{r\to t}(\mathbf{x}) = \mathcal{F}^\top_{t\to r}(\mathbf{x}) \, \mathbf{u}^*(\mathbf{\Phi}_{t\to r}(\mathbf{x}),r)$, and then convert the long-range mapped adjoint velocity to short-range advected adjoint velocity $\mathbf{u}^{*A}_{t'\to t}(\mathbf{x})$, where $t'$ is the last time step of time $t$ (\revise{see Supplementary Section~B.1} for proof):
\begin{equation}\label{eq:long_short_conversion}
\begin{aligned}
    \mathbf{u}^{*A}_{t'\to t}(\mathbf{x}) = \mathbf{u}^{*M}_{r\to t}(\mathbf{x}) + \mathcal{F}^\top_{t\to r}(\mathbf{x}) \revise{\mathbf{\Lambda}}^u_{r\to t'}(\mathbf{\Phi}_{\revise{t\to r}}(\mathbf{x})) + 2\nabla \mathbf{u}^\top \mathbf{u}^*\Delta t\revise{,}
\end{aligned}
\end{equation}
where the advected adjoint velocity $\mathbf{u}^{*A}_{t'\to t}(\mathbf{x})$ is the velocity advected directly by the adjoint advection equation $\left( \frac{\partial}{\partial t} + (\mathbf{u} \cdot \nabla) \right) \mathbf{u}^* =\nabla \mathbf{u}^\top \mathbf{u}^*$  from the previous backward timestep $t'$.  After mapped adjoint passive field $\mathbf{\xi}^{*M}_{r\to t}(\mathbf{x}) = \mathbf{\xi^*}(\mathbf{\Phi}_{t\to r}(\mathbf{x}),r)$ is calculated,  with path integrator $\revise{\mathbf{\Lambda}^\xi_{r\to t'}}$ and current source term $\frac{\partial J}{\partial \xi}$, the current adjoint passive field is updated as:
\begin{equation}\label{eq:adjoint_passive_calculatoin}
    \begin{aligned}
        \mathbf{\xi^*}(\mathbf{x},\revise{t}) = \mathbf{\xi}^{*M}_{r\to t}(\mathbf{x}) +  \mathbf{\Lambda}_{r\to t'}^\xi(\Phi_{t\to r}(\mathbf{x}))- \frac{\partial J}{\partial \xi}(\mathbf{x})\Delta t\revise{.}
    \end{aligned}
\end{equation}

\paragraph{Accumulated Effect} Then we proceed to compute the accumulated contributions from terms other than the mapping.  After the $\xi^*\nabla\xi$ is calculated, together with the viscous term $\nu \Delta \revise{\mathbf{u}^*}$ calculated from $\mathbf{u}^{*A}_{t'\to t}(\mathbf{x})$ and the source term of the objective functional $\frac{\partial J}{\partial \mathbf{u}}$, the unprojected velocity \revise{${\mathbf{u}^*_t}^{\text{up}}{(\mathbf{x})}$} is calculated as

\begin{equation}\label{eq:unprojected_adjoint_velocity}
    \begin{aligned}
        \revise{{\mathbf{u}^{*\text{up}}_t}}{(\mathbf{x})} = \mathbf{u}^{*A}_{t'\to t}(\mathbf{x}) + (\xi^*\nabla \xi -\frac{1}{\rho} \nabla p^* + \nu \Delta \mathbf{u}^* - \frac{\partial J}{\partial \mathbf{u}})\Delta t\revise{.}
    \end{aligned}
\end{equation}

\paragraph{Projection} To obtain the final adjoint velocity at the current timestep, an adjoint Poisson equation \revise{is solved} with the adjoint non-through boundary condition \cite{stuck2012adjoint}:
    \begin{equation}\label{eq:adjoint_Poisson}
        \begin{aligned}
            \frac{\Delta t}{\rho}\Delta p^* = \nabla\cdot {\mathbf{u}^{*\text{\revise{up}}}_{t}}\revise{,}\\
            \mathbf{u}^*_{t} \cdot \revise{\mathbf{n}} = 0, \mathbf{x} \in \partial_b \mathbb{U}_{t} \revise{,}
        \end{aligned}
    \end{equation}
where $\mathbf{n}$ is the normal vector of the solid boundary, and $\partial_b \mathbb{U}_{t}$ denotes the solid boundary of the domain. Then calculate the final adjoint velocity at current time by projection:
\begin{equation}\label{eq:adjoint_projection}
    \begin{aligned}
          \revise{{\mathbf{u}^*_{t}}} = \revise{{\mathbf{u}^{*\text{up}}_{t}}} - \frac{\Delta t}{\rho} \nabla p^* \revise{.}    
    \end{aligned}
\end{equation}

\begin{figure}[t]
    \centering
    \includegraphics[width=1.0\linewidth]{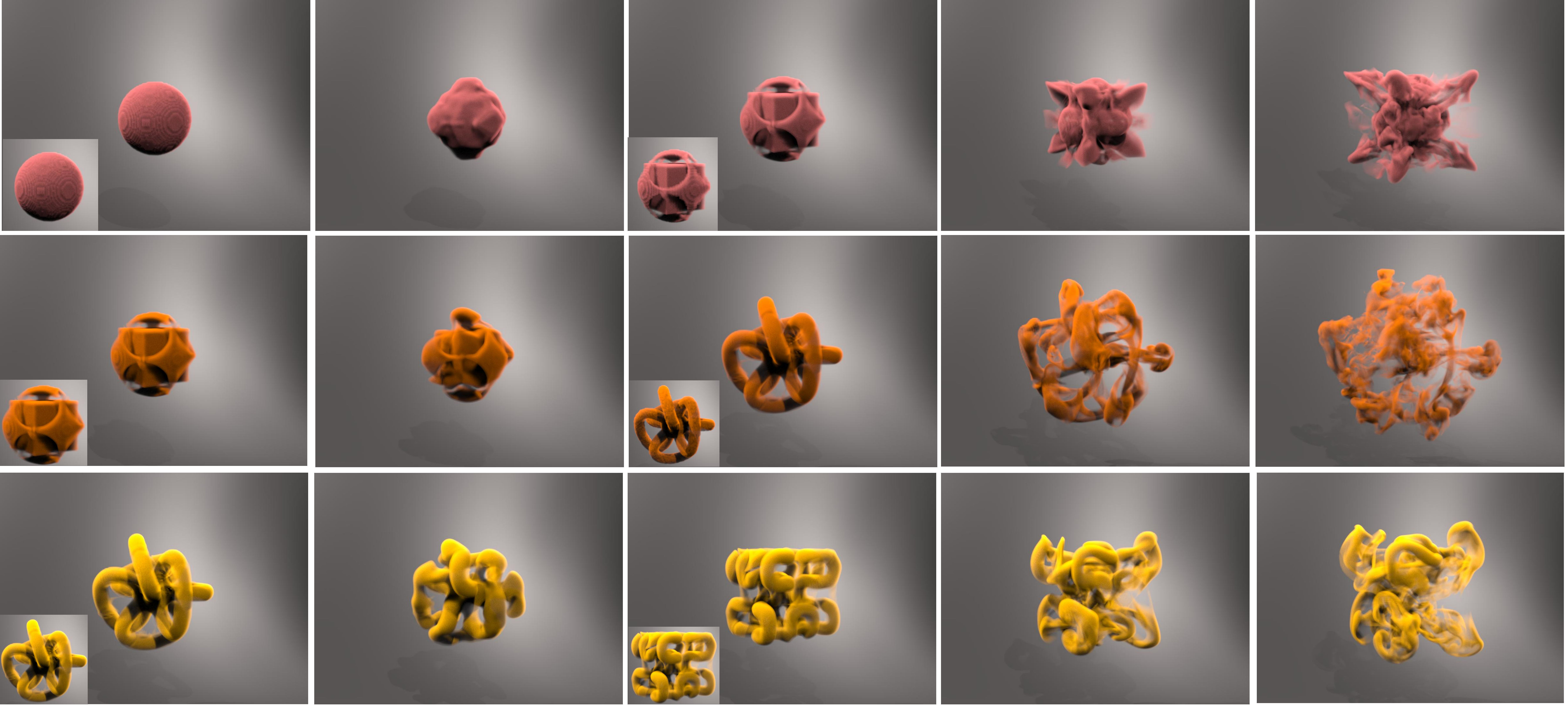}
    \caption{\textbf{3D shape morphing with complex topologies.}  A sequence of 3D shape morphing tasks demonstrating smooth transitions between complex topological structures at frame 0, 50, 100, 150, and 200. Each row illustrates a progressive transformation from a simple to a highly intricate shape, with insets showing the corresponding target geometries. The morphing process preserves topological features while gradually introducing geometric complexity and fine details.}
    \label{fig:3DTopo}
\end{figure}

\paragraph{Path Integrator Update} Subsequently, it is necessary to accumulate the adjoint source term, adjoint viscous term and adjoint pressure gradient $-\nabla p^*$ into the path integrator $\mathbf{\Lambda}^u_{r\to t}$ for long-short term mapping conversion next step and accumulate $\frac{\partial J}{\partial \xi}$ into $\mathbf{\Lambda}^\xi_{r\to t}$. $\revise{\mathbf{\Lambda}}^u_{r\to t}$ and $\mathbf{\Lambda}^\xi_{r\to t}$ are updated by their definition in \autoref{eq:adjoint_velocity_solution_after_integration} with $\Delta t < 0$ respetively as:
\begin{equation}\label{eq:path_integrator_update}
\begin{aligned}
    \mathbf{\Lambda}^u_{r\to t}(\mathbf{x}) &= \mathbf{\Lambda}^u_{r\to t'}(\mathbf{x})+ \Delta t\mathcal{T}_{r\to t}^\top(\mathbf{x})\Big(2\nabla \mathbf{u}^\top \mathbf{u}^* + \xi^*\nabla \xi \\
    &\qquad\qquad\qquad-\frac{1}{\rho} \nabla p + \nu \Delta \mathbf{u}^* - \frac{\partial J}{\partial \mathbf{u}}\Big)(\mathbf{\Psi}_{\revise{r}\to t}(\mathbf{x}),t),\\ 
    \mathbf{\Lambda}_{r\to t}^\xi(\mathbf{x}) &= \mathbf{\Lambda}^\xi_{r\to t'}(\mathbf{x})- \Delta t \frac{\partial J}{\partial \xi}(\mathbf{\Psi}_{\revise{r}\to t}(\mathbf{x}),t). 
\end{aligned}
\end{equation}

\begin{figure}[t]
    \centering
    \includegraphics[width=1.0\linewidth]{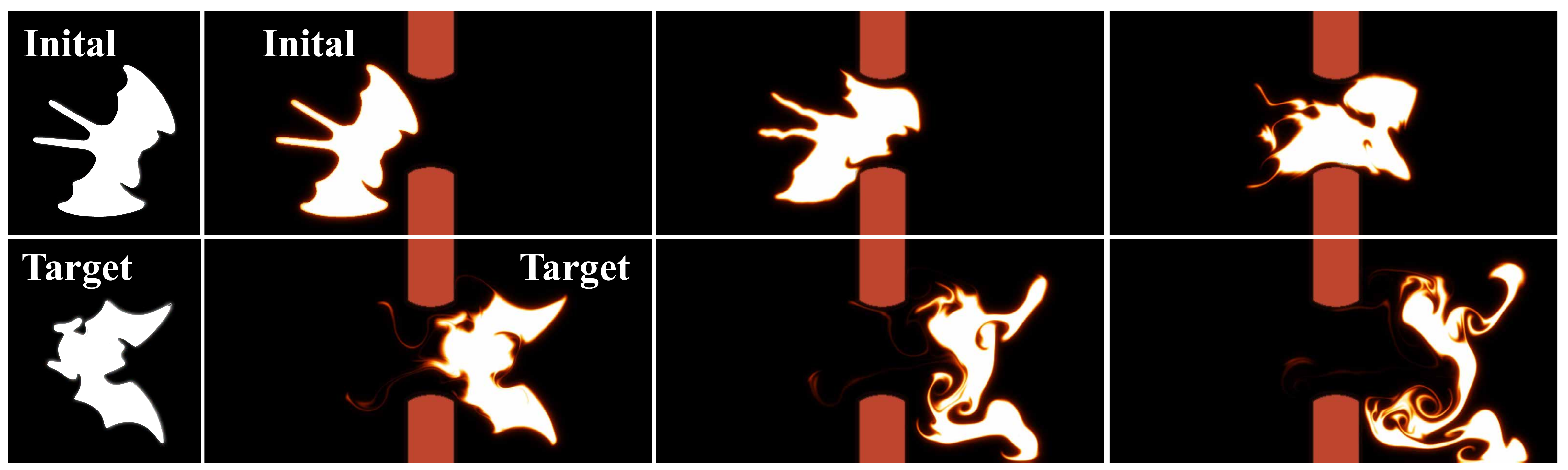}
    \caption{\textbf{Fluid shape morphing with obstacle constraints.} We demonstrate a challenging smoke control task involving obstacle-aware shape matching. By optimizing the control forces, the bat-shaped fluid navigates through the gap and transforms into the target configuration.}
    \label{fig:2Dbat}
\end{figure}

    
\subsection{Long-Short Time-Sparse EFM}\label{sec:Time_Sparse_EFM}
Directly differentiating the flow map forward process poses several challenges. The original flow map methods, including the Eulerian Flow Map (EFM) \cite{deng2023fluid} and Particle Flow Map (PFM) \cite{zhou2024eulerian}, are intensive in terms of both computational cost and memory consumption. Differentiating the forward process requires additional storage for intermediate states, and the 3–5 times backward pass computation further exacerbates the time and memory overhead. However, differentiable simulation requires repeated runs during optimization, making time efficiency critical.
To avoid the expensive $O(m^2)$ flow map evolution in EFM, we adopt time-sparse EFM \cite{Sun_2025} for both the forward \autoref{eq:velocity_solution_after_integration} and backward \autoref{eq:adjoint_velocity_solution_after_integration} computations. In this approach, long-range mapping is applied only at reinit steps, while intermediate steps use semi-Lagrangian for advection and meanwhile accumulate the path integral for later use in flow map calculation at reinit steps.

\begin{figure}[h]
    \centering
    \includegraphics[width=1.0\linewidth]{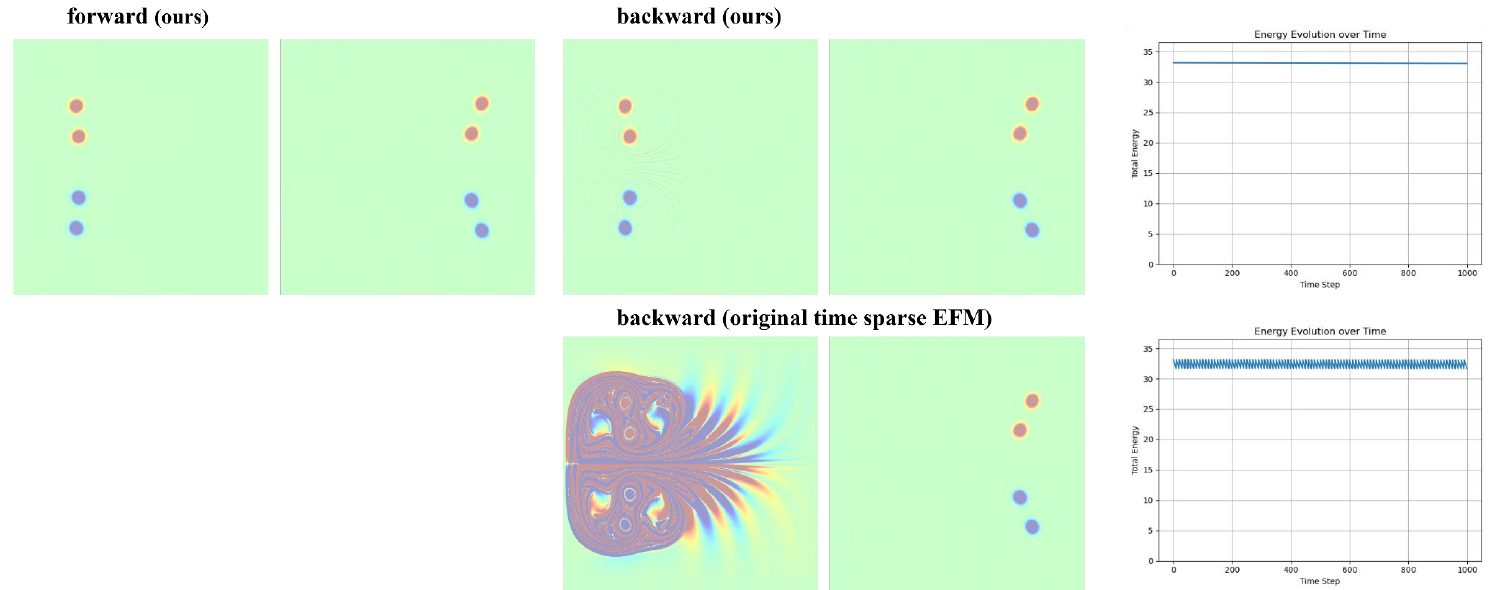}
    \caption{\textbf{Ablation study.} Ablation studies on the leapfrog (left) and single-vortex (right) tests reveal that the original time-sparse EFM fails to compute accurate adjoints: the evolution of leapfrog vortices shows significant errors, while the single-vortex particles exhibit zigzag artifacts in the energy curves, thereby highlighting the necessity of our long-short time-sparse EFM.}
    \label{fig:2Dleapforg_comparison}
\end{figure}

The key idea of time-sparse EFM is to tolerate error accumulation within each reinit interval due to semi-Lagrangian advection, and then correct it at reinit steps using more accurate long-range mapped velocities from the last reinit. As shown in Fig.~\ref{fig:2Dleapforg_comparison} (right), while time-sparse EFM preserves vortices over long timescales, its energy curve shows a sawtooth pattern with clear decay between reinit steps. This is acceptable in the visual effects of computer graphics, where only selected frames are rendered. By aligning the reinit interval with the frame output interval, the visual result remains unaffected.  However, in adjoint calculation in backward pass, this poses a major problem. Unlike visual effects, velocity $\mathbf{u}$ at every step contributes explicitly to the adjoint $\mathbf{u}^*$ through the $\nabla \mathbf{u}^\top \mathbf{u}^*$ term in \autoref{eq:adjoint_NS_equation}. Errors from intermediate steps accumulate via the path integral $\mathbf{\Lambda}_{r\to t}$, significantly degrading the accuracy of $\mathbf{u}^*$—as evidenced by the noticeable artifacts in the adjoint fields of the leapfrogging example shown in Fig.~\ref{fig:2Dleapforg_comparison}.

To address this issue, we propose \textbf{Long-Short Time-Sparse EFM}, an improved version of the original Time-Sparse EFM. We observe that while Time-Sparse EFM maintains accuracy over long temporal distances using long-range flow maps, it suffers from significant errors over short distances. To remedy this, we introduce sparse, short-range flow maps between reinit steps to improve local accuracy.  We call this scheme \textbf{Long-Short Time-Sparse EFM}.  This scheme is used in both forward simulation and backward adjoint calculation.   Here, we take the backward pass as an example. The method for the forward pass is similar \revise{and presented in Algorithm 3 of Supplementary Section A.}

We reinitialize the long-range flow map every $\Delta t_{\text{reinit}}^l = n^{l} \Delta t$ and the short-range flow map every $\Delta t_{\text{reinit}}^s = n^{s} \Delta t$, typically with $n^l = 15 \sim 60$ and $n^s = 1\sim 3$.  \revise{In Algorithm \autoref{alg:general_flowmap}}, we illustrate one long-range reinit cycle starting from the previous long-range reinit time $r'$. Let $r''$ denote the most recent short-range reinit \revise{time}, initialized as $r'' = r'$. We maintain two sets of flow maps, $\mathbf{\Psi}^l_{r'\to t}$, $\mathcal{T}^l_{r'\to t}$, $\mathbf{\Phi}^l_{t_c\to t}$, $\mathcal{F}^l_{t_c\to t}$ for long-range, $\mathbf{\Psi}^s_{r''\to t}$, $\mathcal{T}^s_{r''\to t}$, $\mathbf{\Phi}^s_{t_c\to t}$, $\mathcal{F}^s_{t_c\to t}$ for short-range and two path integrators $\mathbf{\Lambda}_{r'\to t}^{u,l}$, $\mathbf{\Lambda}_{r'\to t}^{\xi,l}$ and $\mathbf{\Lambda}_{r''\to t}^{u,s}$, $\mathbf{\Lambda}_{r''\to t}^{\xi,s}$ for long-range and short-range respectively, where $t_c < r'$ is the current time \revise{and $t$ serves as the evolving time variable during integration}.

\begin{algorithm}[t]
\caption{\revise{Long-Short Time-Sparse EFM of Backward Pass}}
\label{alg:general_flowmap}
\begin{flushleft}
        \revise{\textbf{Initialize:} short-range
reinit time $r''$ to long-range reinit time $r'$. }
\end{flushleft}
\begin{algorithmic}[1]
  \For{each time step $t_c$ between $r'$ and $r'+\Delta t_{\text{reinit}}^l$,}
    \State March $\mathbf{\Psi}^l_{r'\to t_c}$,$\mathbf{\Psi}^s_{r''\to t_c}$,$\mathcal{T}^l_{r'\to t_c}$,$\mathcal{T}^s_{r''\to t_c}$ one step\revise{;} \hfill $\triangleright$ eq.~\ref{eq:nfm_evolution2} 
  \If{$\exists m\in \mathbb{Z}$, st. $t_c = \revise{r'}+m\Delta t_{\text{reinit}}^s$}
  \State \highlight{Calculate $\mathbf{\Phi}^s_{t_c\to r''}$ and $\mathcal{F}^s_{t_c\to r''}$ by integrating $\mathbf{\Phi}^s_{t_c\to t}$ and} \\$\qquad\qquad$ \highlight{$\mathcal{F}^s_{t_c\to t}$ from $t_c$ to $r''$}\revise{;} \hfill $\triangleright$ eq.~\ref{eq:Phi_F_advection}
  \State Do mapping and conversion with $\mathbf{\Phi}^s_{t_c\to r''}$ and \\$\qquad\qquad$ $\mathcal{F}^s_{t_c\to r''}$ \revise{;}  \hfill $\triangleright$ \revise{eq.~\ref{eq:long_short_conversion}}
  \State Reset $\mathbf{\Psi}^s_{r''\to t_c}$ and $\mathcal{T}^s_{r''\to t_c}$ \revise{and set $r''= t_c$}\revise{;}
  \Else
  \State Calculate advection with semi-\revise{Lagrangian} method\revise{;}
  \EndIf
  \State  Calculate source term and projection\revise{;}
  \State  Update $\mathbf{\Lambda}_{r'\to t_c}^{u,l}$, $\mathbf{\Lambda}_{r'\to t_c}^{\xi,l}$ and $\mathbf{\Lambda}_{r''\to t_c}^{u,s}$, $\mathbf{\Lambda}_{r''\to t_c}^{\xi,s}$\revise{;} \hfill $\triangleright$ eq.~\ref{eq:path_integrator_update}
  \If{$t_c = r'+\revise{\Delta t_{\text{reinit}}^l}$}
    \State  Calculate $\mathbf{\Phi}^l_{t_c\to r'}$ and $\mathcal{F}^l_{t_c\to r'}$ by integrating $\mathbf{\Phi}^l_{t_c\to t}$ and \\$\qquad\qquad$ $\mathcal{F}^l_{t_c\to t}$ from $t_c$ to $r'$\revise{;} \hfill $\triangleright$ eq.~\ref{eq:Phi_F_advection}
    \State Correct results by adding long-range mapping with \\$\qquad\qquad$ $\mathbf{\Phi}^l_{t_c\to r'}$, $\mathcal{F}^l_{t_c\to r'}$ and path integrators.  \hfill $\triangleright$ \revise{eq.~\ref{eq:adjoint_velocity_solution_after_integration}}
  \EndIf
  \EndFor{}
\end{algorithmic}
\end{algorithm}

In Algorithm \autoref{alg:general_flowmap}, the blue-highlighted parts indicate the components added by Long-Short Time-Sparse EFM compared to Time-Sparse EFM, corresponding to the blue lines in Fig.~\ref{fig:different_mapping} (c)(d). Our method retains the $O(m)$ time complexity of flow map evolution, with a one-time overhead in the number of evolution steps.  As in the original Time-Sparse EFM, we use long-range flow maps to maintain accuracy over large time intervals, while short-range flow maps ensure accuracy between sparse long-range updates, preventing error accumulation of $\mathbf{u}^*$ caused by inaccurate intermediate velocities. 

\section{Numerical Algorithm}\label{sec:numerical}
We use a grid $\mathbb{G}$ with spacing $\Delta x$, where $\mathbf{x}_g$ denotes the position of grid point $g \in \mathbb{G}$. We denote the field value at grid point $g$ by the subscript $_g$. With a fixed time step $\Delta t$, we define $t_i =i\Delta t$ for $i = 0, \ldots, n$, where $t_0 = s$ and $t_n = r$.  \revise{Our complete algorithm consists of a forward pass for computing physical quantities and a backward pass for computing their adjoints. Algorithm~\autoref{alg:simulation_scheme} illustrates the backward pass, while the forward pass is presented in Algorithm 4 of Supplementary Section C, as forward pass calculation is not the main focus of this paper.}

\begin{algorithm}[t]
\caption{Long-Short Time-Sparse EFM for \revise{A}djoint \revise{F}ield}
\label{alg:simulation_scheme}
\begin{flushleft}
        \textbf{Initialize:} $\mathbf{u}^*_{\revise{r''}, g}$, $\mathbf{u}^*_{r',g}$ to initial adjoint velocity; $\mathcal{T}^s_{\revise{r''} \to \revise{t_c},g}$, $\mathcal{F}^s_{\revise{t_c} \to \revise{r''},g}$, $\mathcal{T}^l_{r' \to t_c,g}$, $\mathcal{F}^l_{t_c \to r',g}$ to $\mathbf{I}$; $\mathbf{\Psi}^s_{\revise{r''} \to \revise{t_c},g}$, $\mathbf{\Phi}^s_{\revise{t_c} \to \revise{r''},g}$, $\mathbf{\Psi}^l_{r' \to t_c,g}$, $\mathbf{\Phi}^l_{t_c \to r',g}$ to $\mathbf{x}_g$; $\revise{\mathbf{\Lambda}}_{r'\to \revise{t_c},g}^{u,l}$, $\revise{\mathbf{\Lambda}}_{r'\to \revise{t_c},g}^{\xi,l}$ and $\revise{\mathbf{\Lambda}}_{r''\to \revise{t_c},g}^{u,s}$, $\revise{\mathbf{\Lambda}}_{r''\to \revise{t_c},g}^{\xi,s}$ to 0; $r'$, $r''$, $t_c$ to $r$.
\end{flushleft}
\begin{algorithmic}[1]
\For{each time step $t_c$ from $t_n$ to $t_0$}
\State Load midpoint velocity $\mathbf{u}_g^\text{mid}$; 
\State March \revise{$\mathbf{\Psi}^l_{r'\to t_c,g}$,$\mathbf{\Psi}^s_{r''\to t_c,g}$,$\mathcal{T}^l_{r'\to t_c,g}$,$\mathcal{T}^s_{r''\to t_c,g}$}\revise{ one step;} \hfill $\triangleright$ eq.~\ref{eq:nfm_evolution2}
\If{$c \Mod {n^s} = 0$}
\State Integrate $\mathcal{F}^s_{\revise{t_c} \to r'',g}$ and $\mathbf{\Phi}^s_{ \revise{t_c} \to r'',g}$ from $t_c$ to $r''$\revise{;}\hfill $\triangleright$ eq. \ref{eq:Phi_F_advection}
\State Calculate mapped adjoint velocity $\mathbf{u}^{*M}_{r''\to t_c\revise{,g}}$ and convert to one-step advected adjoint velocity $\mathbf{u}^{*A}_{t_{c+1}\to t_c\revise{,g}}$\revise{;} \hfill $\triangleright$ eq. \ref{eq:long_short_conversion}
\State Set initial time for short mapping $r''$ to \revise{$t_c$};
\State Reinitialize $\mathcal{T}^{\revise{s}}_{r''\to t_c,g}$ to $\mathbf{I}$, $\mathbf{\Psi}^{\revise{s}}_{r''\to t_c,g}$ to $\mathbf{x}_g$;
\State Calculate $\xi^*_{t_{c},g}$ by $\mathbf{\Phi}^s_{ \revise{t_c} \to r'',g}$ and $\mathbf{\Lambda}_{\revise{r''}\to t_{c+1},g}^{\xi,{\revise{s}}}$\revise{;} \hfill $\triangleright$ eq. \ref{eq:adjoint_passive_calculatoin}
\Else
\State Calculate $\mathbf{u}^{*A}_{t_{c+1}\to t_{c},g}$ and $\xi^{*A}_{t_{c+1}\to t_{c},g}$ by \revise{semi-Lagrangian};
\State Calculate  $\xi^*_{t_{c},g}=\xi^{*A}_{t_{c+1}\to t_{c},g}-\Delta t\frac{\partial J}{\partial \xi}$\revise{;} 
\EndIf
\State Calculate adjoint viscous term $[\nu \Delta \mathbf{u}^*]_g$\revise{;} \hfill $\triangleright$ eq. \ref{eq:viscous_term_calculation}
\State Calculate coupling term $[\xi^*\nabla \xi]_g$ and $[\nabla \mathbf{u}^\top \mathbf{u}^*]_g$; \hfill $\triangleright$ eq. \ref{eq:derivative_calculation}
\State Compute source term from objective functional $\frac{\partial J}{\partial \mathbf{u}}$\revise{;}

\State Compute unprojected ${\mathbf{u}_{t_c,g}^{*\text{\revise{up}}}}$; \hfill $\triangleright$ eq. \ref{eq:unprojected_adjoint_velocity}

\State Calculate ${\mathbf{u}_{t_c,g}^*}$ using $p^*$ from \revise{P}ossion equation; \hfill $\triangleright$ eq. \ref{eq:adjoint_Poisson},\ref{eq:adjoint_projection}

\State Update both short and long path integrator $\mathbf{\Lambda}_{r'\to t_c,g}^{u,l}$, $\mathbf{\Lambda}_{r'\to t_c,g}^{\xi,l}$ and $\mathbf{\Lambda}_{r''\to t_c,g}^{u,s}$, $\mathbf{\Lambda}_{r''\to t_c,g}^{\xi,s}$\revise{;} \hfill $\triangleright$ eq.~\ref{eq:path_integrator_update}

\If{$c \Mod {n^l} = 0$}
    \State Integrate $\mathcal{F}^l_{\revise{t_c} \to r',g}$ and $\mathbf{\Phi}^l_{ \revise{t_c} \to r',g}$ from $t_c$ to $r'$\revise{;}\hfill $\triangleright$ eq. \ref{eq:Phi_F_advection}
    \State Calculate accurate $\xi_{t_c,g}^*$ and $\mathbf{u}_{t_c,g}^*$ by mapping with  $\mathcal{F}^l_{\revise{t_c} \to r',g}$, $\mathbf{\Phi}^l_{ \revise{t_c} \to r',g}$ and integrator $\mathbf{\Lambda}_{r'\to t_c,g}^{u,l}$, $\mathbf{\Lambda}_{\revise{r'}\to t_c,g}^{\revise{\xi,l}}$\revise{;} \hfill $\triangleright$ eq.~\ref{eq:adjoint_velocity_solution_after_integration}
    \State Set initial time for long mapping $r'$ to \revise{$t_c$}\revise{.}
\EndIf
\EndFor{}
\end{algorithmic}
\end{algorithm}

\paragraph{Interpolation}To compute the mapping, we interpolate the mapped field using a second-order kernel with a support radius of $1.5\Delta x$. For example, the computation of $\mathbf{u}^{*M}_{r' \to t_c}(\mathbf{x}) = \mathcal{F}_{t_c\to r'}(\mathbf{x})\mathbf{u}^*(\mathbf{\Phi}_{t_c\to r'}(\mathbf{x}\revise{),r'})$ is given by:
\begin{equation}
    \begin{aligned}
        \mathbf{u}^{*M}_{r' \to t_c,g}= \mathcal{F}_{t_c\to r',g}\sum_{g'\in N(\mathbf{\Phi}_{t_c\to r',g})}\mathbf{u}^*_{r', g'}w(\mathbf{x}_{g'}-\mathbf{\Phi}_{t_c\to r',g})\revise{,}
    \end{aligned}
\end{equation}
where $N(\mathbf{\Phi}_{t_c \to r',g}) = \{ g' \mid w(\mathbf{x}_{g'} - \mathbf{\Phi}_{t_c \to r',g}) > 0 \}$ denotes the set of grid points neighboring $\mathbf{\Phi}_{t_c \to r',g}$.

\paragraph{Midpoint Velocity} Following \cite{deng2023fluid}, the flow maps and their Jacobians, $\mathcal{F}$ and $\mathcal{T}$, are advected using the fourth-order Runge-Kutta method to solve \autoref{eq:Phi_F_advection} and \autoref{eq:nfm_evolution2}, respectively. The midpoint velocity $\mathbf{u}_r^{\text{mid}}$ is computed according to Algorithm 2 in \cite{deng2023fluid} by marching the velocity field forward by half a time step. Since these midpoint velocities are also required during the backward pass for evolving the flow maps and solving \autoref{eq:adjoint_velocity_solution_after_integration}, we store them at each step to avoid redundant computation. Unlike those autodiff methods, which require retaining all intermediate velocities in the computation graph, our method allows storing them on disk or in memory without using GPU.

\paragraph{BFECC for Mapping}
Since the flow maps we evolve are bidirectional, following \cite{Sun_2025,deng2023fluid}, we apply back and forth error compensation and correction (BFECC) during the mapping process using the flow maps. Taking the mapping of adjoint velocity $\mathbf{u}^*_{r'\to t_c}(\mathbf{x}) = \mathcal{F}_{t_c\to r'}(\mathbf{x})\mathbf{u}^*(\mathbf{\Phi}_{t_c\to r'}(\mathbf{x}),r')$ as an example, the BFECC procedure is as follows:
\begin{equation}\label{eq:BFECC}
    \begin{aligned}
        \mathbf{u}^{*(1)}_{r'\to t_c}(\mathbf{x}) &= \mathcal{F}_{t_c\to r'}(\mathbf{x})\mathbf{u}^*(\mathbf{\Phi}_{t_c\to r'}(\mathbf{x}),r') \revise{,}\\
        \mathbf{u}^{*(2)}_{t_c\to r'}(\mathbf{x}) &= \mathcal{T}_{r'\to t_c}(\mathbf{x})\mathbf{u}^{*(1)}_{r'\to t_c}(\mathbf{\Psi}_{\revise{r'\to t_c}}(\mathbf{x})) \revise{,}\\
        \mathbf{u}^*_{r'\to t_c}(\mathbf{x}) &= \mathbf{u}^{*(1)}_{r'\to t_c}(\mathbf{x}) + \frac{1}{2}\mathcal{F}_{t_c\to r'}(\mathbf{x})(\mathbf{u}_{\revise{r'}}^*- \mathbf{u}^{*(2)}_{t_c\to r'}\revise{)}(\mathbf{\Phi}_{t_c\to r'}(\mathbf{x}))\revise{,}
    \end{aligned}
\end{equation}
where the superscripts $^{(1)}$ and $^{(2)}$ denote intermediate steps in the BFECC process.  Other terms, such as $\mathcal{F}_{t_c\to r'}(\mathbf{x})\Lambda^u_{r'\to t_c}(\mathbf{\Phi}_{t_c\to r'}(\mathbf{x}))$, which rely on flow maps for calculation, are similarly computed using this strategy.

\paragraph{Pressure Projection}  For 2D simulations, we solve the Poisson equation using a multigrid preconditioned conjugate gradient solver (MGPCG) as in \cite{zhou2024eulerian,deng2023fluid}. For 3D simulations, to improve computational efficiency, we adopt a fast matrix-free algebraic multigrid preconditioned conjugate gradient (AMGPCG) solver as used in \cite{Sun_2025}.    

\begin{figure*}[t]
    \centering
    \hspace{-0.5cm}
    \includegraphics[width=0.24\linewidth]{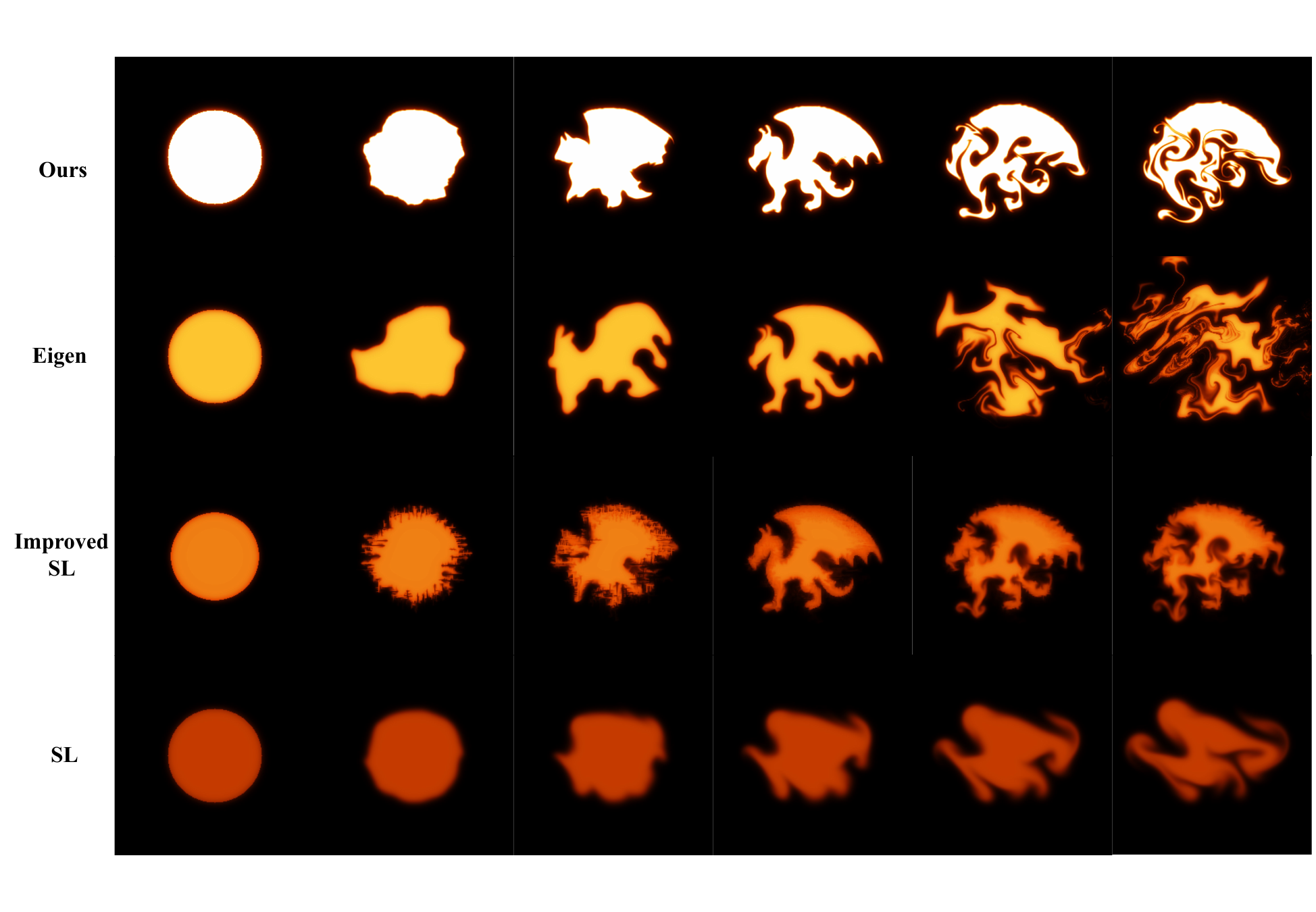}
    \includegraphics[width=0.24\linewidth]{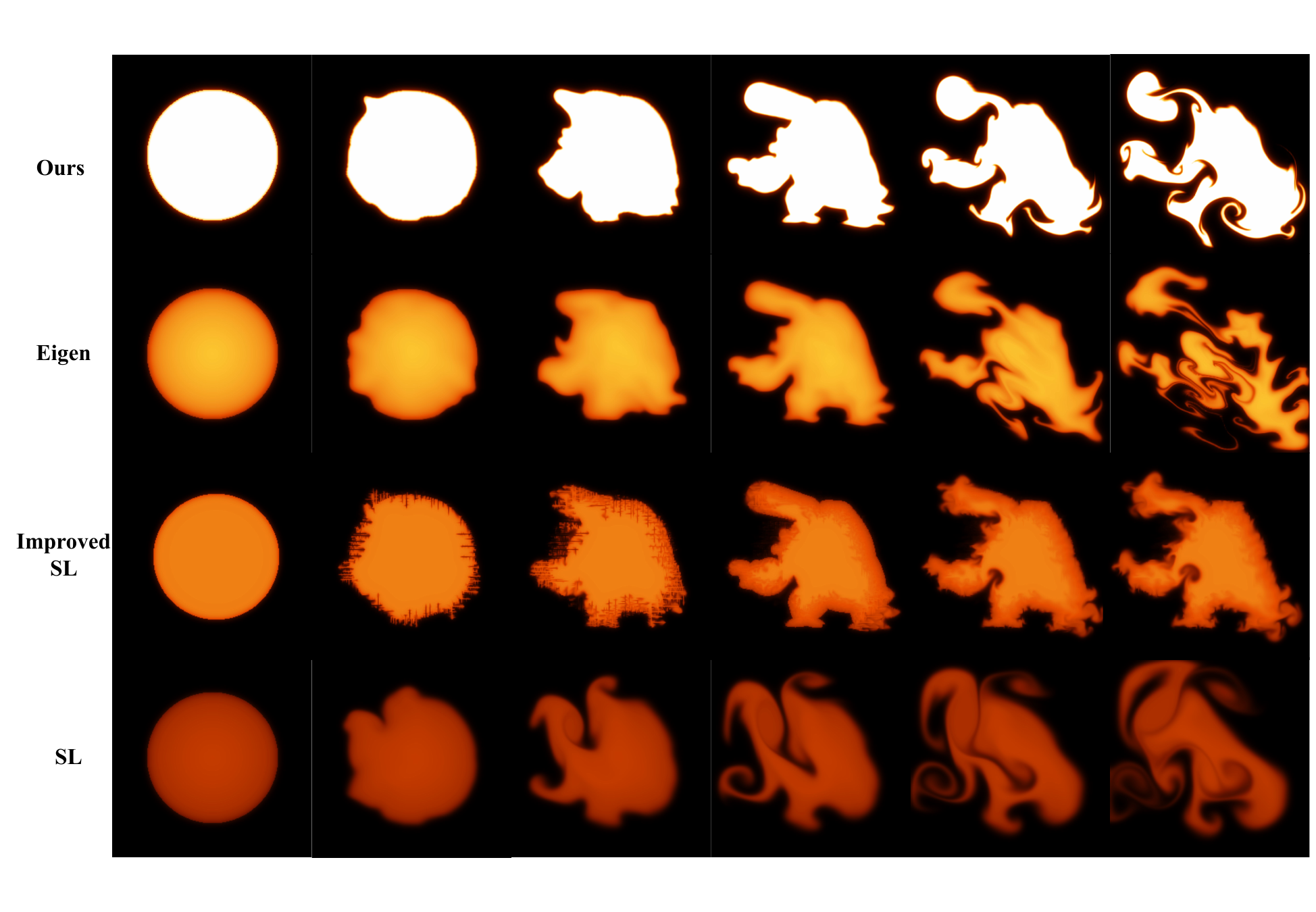}
    \includegraphics[width=0.24\linewidth]{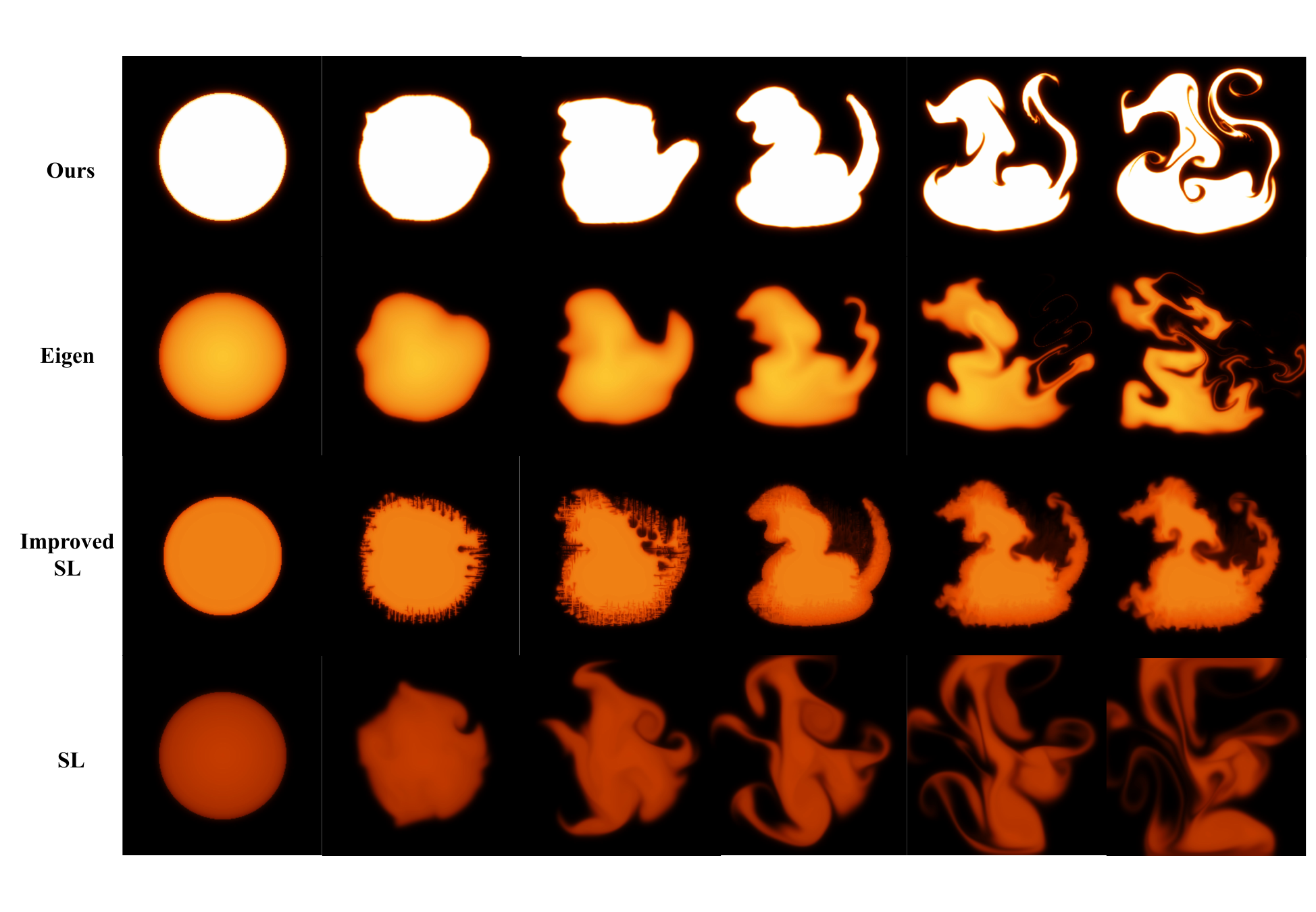}
    \includegraphics[width=0.28\linewidth]{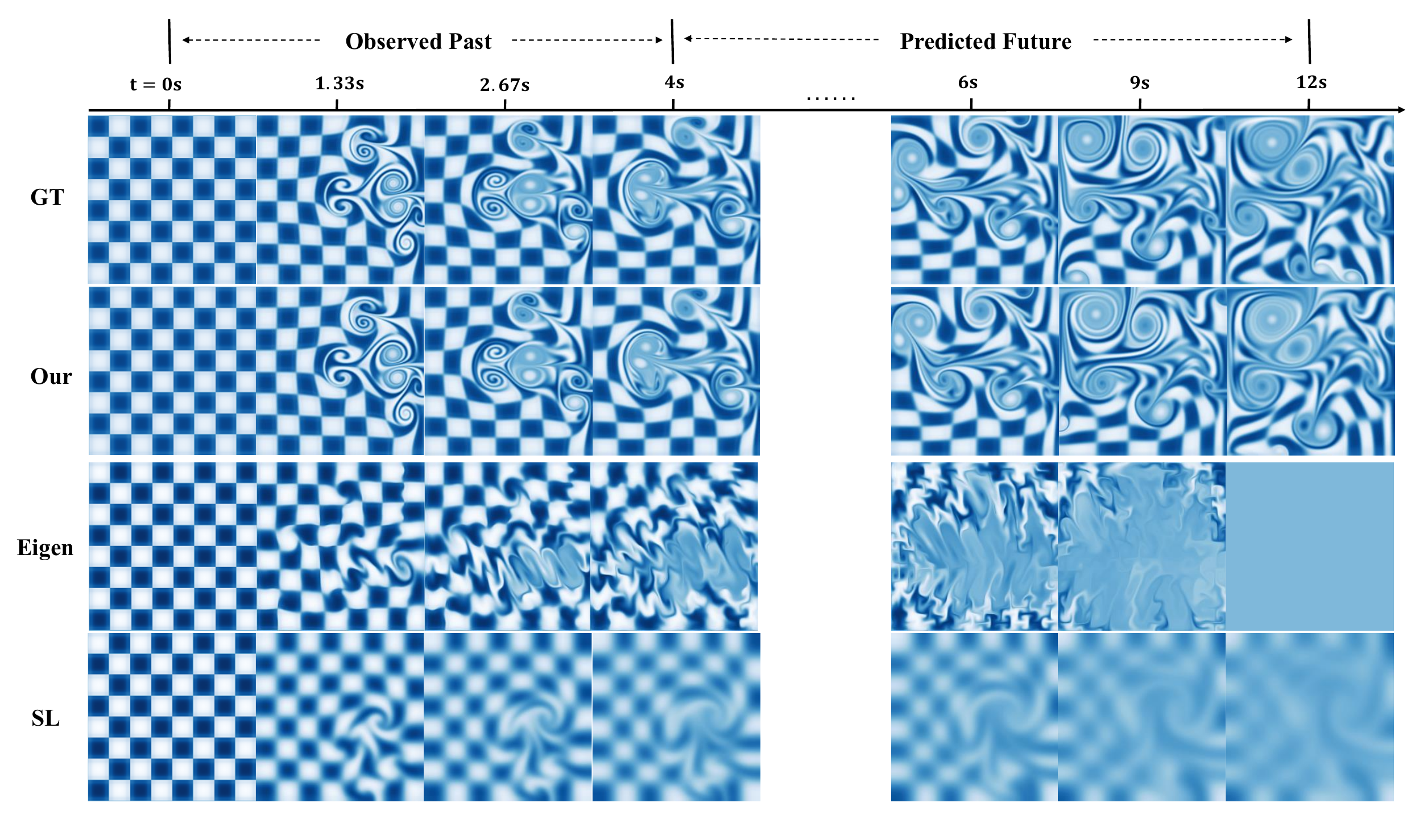}
    \hspace{-0.5cm}
    \caption{\textbf{Comparison with different differentiable solvers.} We compare our method with EigenFluids and semi-Lagrangian (SL) methods on two tasks: 2D smoke control and vortex inference. 
    \textit{Columns 1–3:} 2D smoke control tasks (Dragon, Turtle, Snake). 
    SL fails to achieve precise fluid control, while EigenFluids and SL with optimized control forces (Improved SL) can guide the fluid but suffer from poor advection, leading to unrealistic motion. 
    In contrast, our approach achieves accurate control while preserving fine details and overall volume, resulting in more realistic fluid behavior. 
    \textit{Column 4:} Vortex inference task, where only our method succeeds, highlighting its superior accuracy.}

    \label{fig:2DShapeComparisons}
\end{figure*}


\paragraph{Derivatives Calculation} On the grid, we compute the viscous term using second-order finite difference schemes, as
\begin{equation}\label{eq:viscous_term_calculation}
    \begin{aligned}
        [\nu\Delta \mathbf{u}]_g = \nu \sum_{g'\in N_g}(\mathbf{u_g'} - \mathbf{u_g})/\revise{(|N_g|\Delta x^2)}\revise{,}        
    \end{aligned}
\end{equation}
where $N_g$ represents the adjacent grid points of $g$ with $|N_g| =4$ for 2D and $|N_g| =6$ for 3D.  For the coupling terms $\nabla \mathbf{u}^\top \mathbf{u}^*$ and $\mathbf{\xi}^* \nabla \mathbf{\xi}$ in \autoref{eq:long_short_conversion} and \autoref{eq:unprojected_adjoint_velocity}, we use the third-order kernel-based interpolation to calculate as
\begin{equation}\label{eq:derivative_calculation}
    \begin{aligned}
    [\nabla \zeta]_g = \sum_{g'\in N_g^{\revise{w_3}}} \zeta_g \nabla_{\revise{x_g}} w_3(\revise{\mathbf{x}}_g - \revise{\mathbf{x}}_{g'})\revise{,}
    \end{aligned}
\end{equation}
where $\zeta$ is a field (either $\mathbf{u}$ or $\xi$), and $\revise{w_3}$ is a third-order kernel with compact support of radius $2\Delta x$. The neighbor set $N^{w_3}_g = \{ g' \mid w_3(\revise{\mathbf{x}}_{g'} - \revise{\mathbf{x}}_g) > 0 \}$ includes grid points within the kernel support of $g$.  When computing $\nabla \mathbf{u}^\top \mathbf{u}^*$, for consistency with $\mathbf{u}^*$ , $\mathbf{u}$ used in $\nabla \mathbf{u}^\top \mathbf{u}^*$ is calculated backward along with $\mathbf{u}^*$.

\section{Results and Discussion}\label{sec:examples}

\revise{In this section, we first validate the accuracy of our method on specific flow fields and objective functionals whose adjoint velocities admit analytical solutions. We then demonstrate our approach on three representative tasks: vortex dynamics inference from videos, vortex control, and smoke control. For these tasks, we compare our method with prior differentiable solvers, highlighting both quantitative and qualitative improvements. For each task, we will present the objective functional integrand $J$ (see \autoref{eq:object_function_def}) along with experiment-specific optimization parameters.}

\begin{figure}[h]
    \centering
    \includegraphics[width=1.0\linewidth]{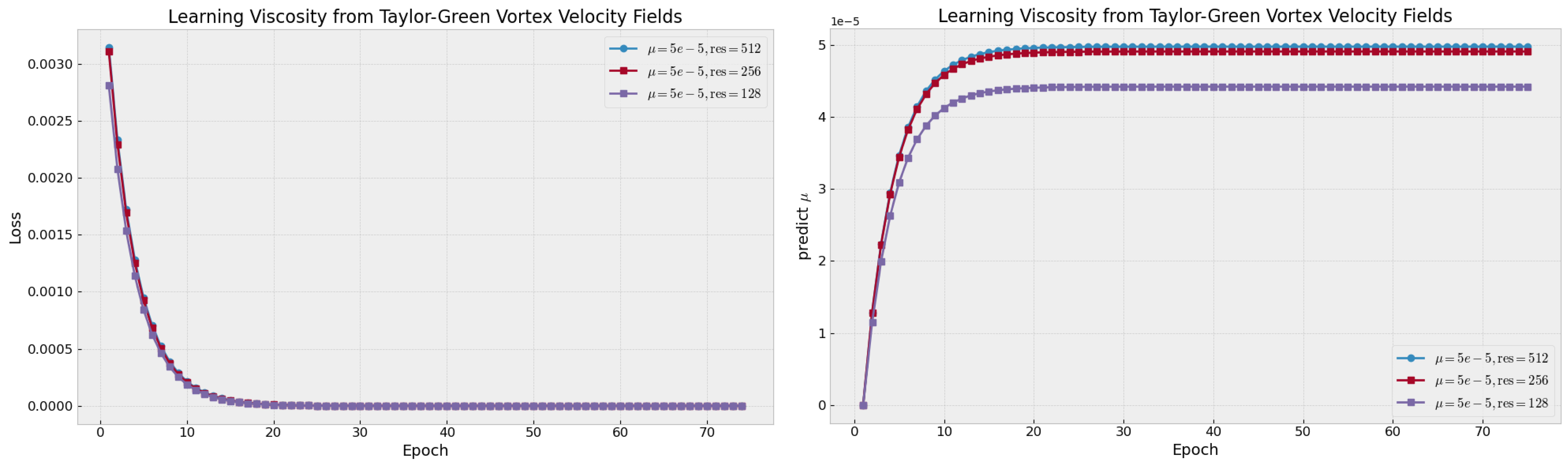}
    \caption{\textbf{Taylor-Green velocity field viscous inference.} We use the analytical solution as the target to test if our method can infer the viscosity from the velocity field. Results at resolutions 128, 256, and 512 are shown.}
    \label{fig:taylor_plot}
\end{figure}

\begin{figure}[h]
    \centering
    \includegraphics[width=1.0\linewidth]{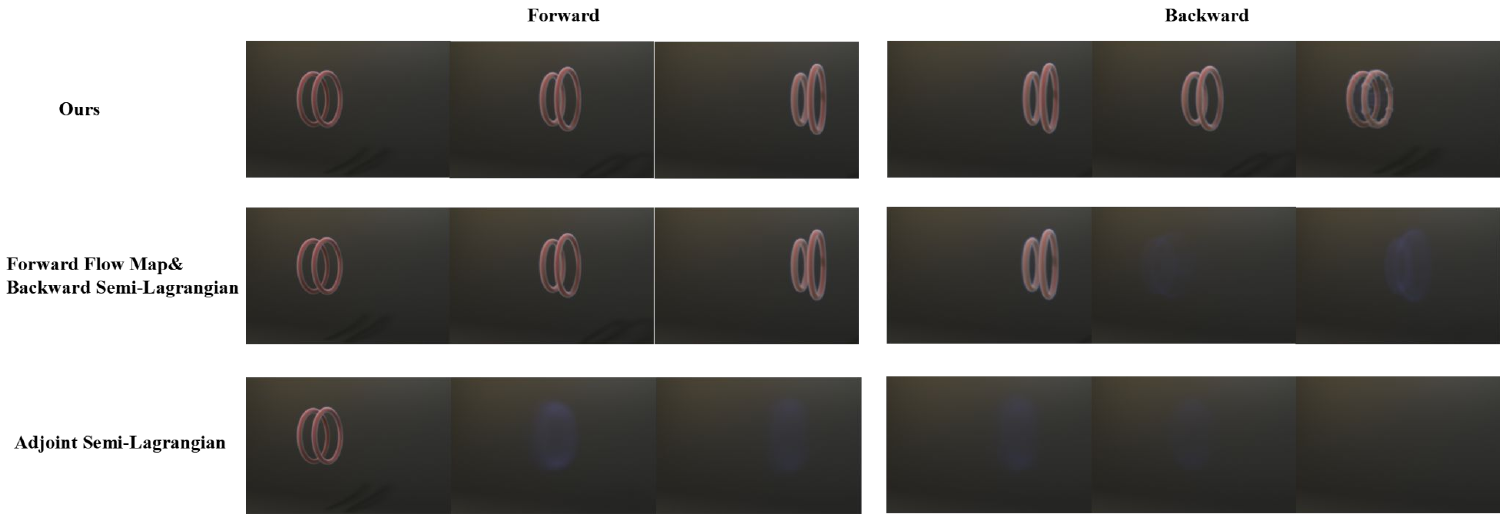}
    \caption{\textbf{3D leapfrog comparison of forward and backward processes using different methods.}
    We test replacing both the forward and backward computations with semi-Lagrangian methods, and the results confirm that flow map methods are essential for both directions. This experiment also demonstrates the accuracy of our method in computing the adjoint.}
    \label{fig:3D_leapforg}
\end{figure}

\paragraph{Validation \& Ablation Test} First, \revise{we} validate on cases with known adjoints for correctness of adjoint calculation: for incompressible flow with rigid walls, choosing
$J=\frac{1}{2}\delta(t-r)\|\mathbf{u}\|^2$ forces $\mathbf{u}^*_t=\mathbf{u}_t$ ($r\!\ge\!t\!\ge\!s$), giving a direct velocity check. \revise{For all experiments, we set the simulation to run for 100-500 steps and performed 200–800 optimization iterations. }\revise{Fig.~\ref{fig:2Dleapforg_comparison} confirms the correctness of our results for single one vortex and leapfrog vortices}, and also shows that using Time‑Sparse EFM alone incurs large errors, highlighting the need for our Long‑Short Time‑Sparse EFM. \textbf{3D leapfrog.} Fig.~\ref{fig:3D_leapforg} confirms our adjoint computation for 3‑D leap‑frogging vortex rings— a case standard autodiff fails. It also shows that the semi‑Lagrangian method for forward or backward pass yields incorrect results.  Next, we verify optimization capability through a known analytic flow. \textbf{Viscosity Coefficient Inferring.} On the domain $[0,2\pi]^2$, the Navier–Stokes equations admit the Taylor‑Green vortex $\mathbf{u} = (\cos x \sin y,-\sin x \cos y)e^{-2\nu t}$. Using its analytic velocity at $800\Delta t$ as the target $\mathbf{u}_{\text{target}}$ and treating the viscosity $\nu$ as the optimization parameter, we infer the viscosity by minimizing $\revise{J=\tfrac12\,\delta(t-r)\bigl(\|\mathbf{u}_{\text{target}}\|^{2}-\|\mathbf{u}\|^{2}\bigr)}$ as shown in Fig.~\ref{fig:taylor_plot}.

\begin{figure}[t]
    \centering
    \includegraphics[width=1.0\linewidth]{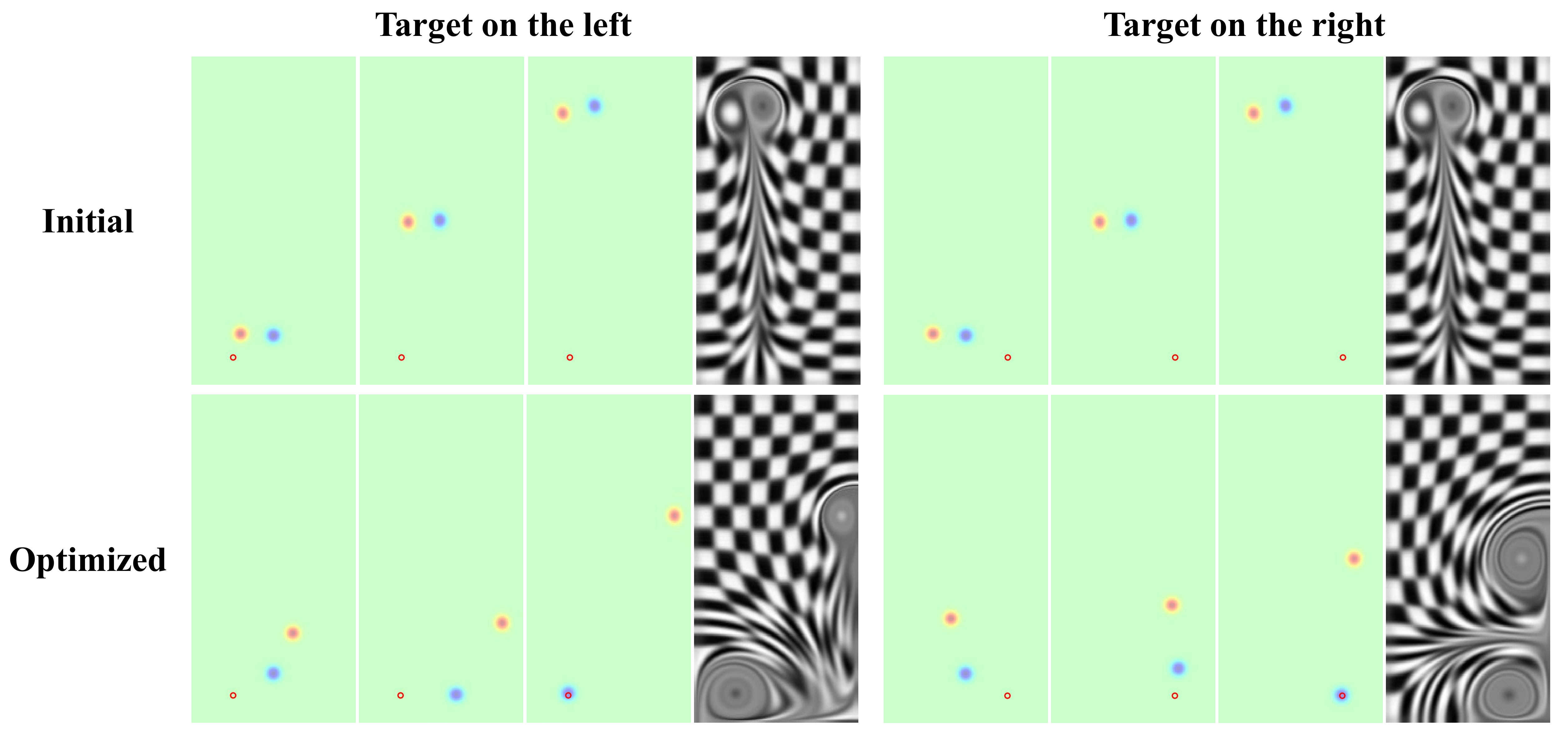}
    \caption{\textbf{Single vortex control via position and vorticity optimization.}  We optimize the position and vorticity of the red vortex to guide the blue vortex toward the target (indicated by the hollow circle). The top row shows results without optimization, while the bottom row shows optimized results that successfully steer the blue vortex to the target at frames 0, 400, and 800, with a checkerboard pattern at frame 800.}
    \label{fig:vortex_control_ex12}
\end{figure}

\begin{figure}[t]
    \centering
    \includegraphics[width=1.0\linewidth]{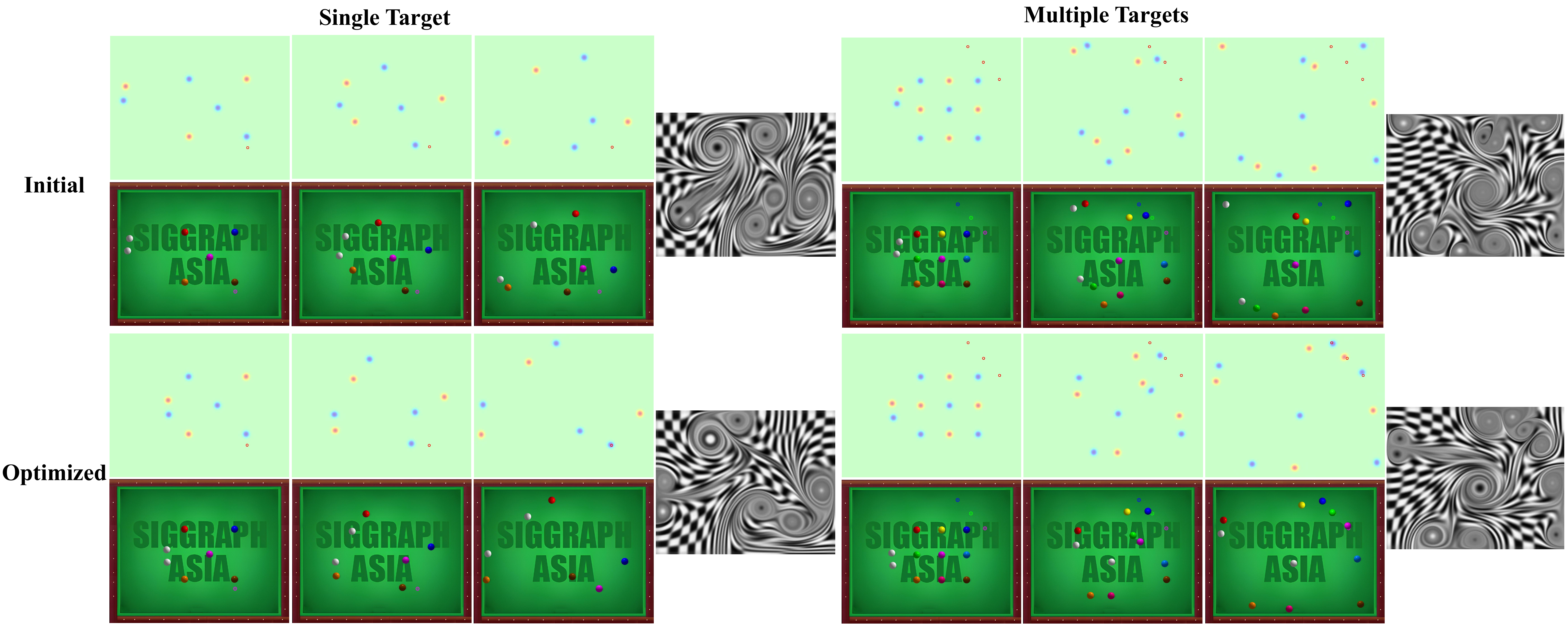}
    \caption{\textbf{Multiple vortex control.}   We optimize the positions and vorticities of the white spheres to accomplish complex control tasks involving single and multiple targets. Hollow circles denote the target positions. The top row shows results with randomly initialized white spheres, while the bottom row displays our optimized outcomes. Snapshots of vorticity and rendered results are provided at frames 0, 400, and 800, with the checkerboard pattern shown at frame 800.}
    \label{fig:vortex_control_ex34}
\end{figure}

\paragraph{Examples} \textbf{(1) Vortex Dynamics Inference from \revise{Videos}.} This task infers an initial velocity field, represented by 16 vortices, from a single RGB video of fluid motion (synthetic or real) \cite{deng2023learning}. The optimization variables are $(c_{x,i}, c_{y,i}, w_i, r_i), i= 1,...,16$ which denote the position, strength, and radius of each vortex respectively. The objective minimizes the difference between simulated frames \revise{$\xi_{i}$} and target video frames \revise{$\xi_{i,\text{gt}}$}: 
\revise{$J = \frac{1}{2}\sum_{k} \sum_{i=1}^3 \delta(t-t_k)|\xi_{i}(\mathbf{x},t) - \xi_{i,\text{gt}}(\mathbf{x},t)|^2$}, where $\xi_{i},i=1,2,3$ represents the passive fields for the R, G, and B channels, \revise{and $t_k$ denotes the time corresponding to the $k$-th frame.}  Optimization is initialized with 16 randomly placed vortices, and the RGB video is the only input of \revise{the optimization process}.  We show some beautiful results of infering \revise{initial} vorticity field from warped \textbf{2D Logo} and \textbf{2D Gradient Background} (Fig.~\ref{fig:2DLogo_and_Gradient}) \revise{examples}, \revise{and} more complex scenarios with obstacles such as \textbf{2D Three Cylinders} and \textbf{2D Leaf Obstacle} (Fig.~\ref{fig:2Dobstacal})\revise{. All} demonstrate the robustness and generality of our method in complex environments, including multiple obstacles and real environment. 
 This task requires strong vortex preservation, which can only be achieved using differentiable flow maps.  
 \textbf{(2) 2D Vortex Control.}  This task optimizes \revise{the initial positions $\mathbf{c}_{j,\text{init}}^{\text{ctrl}}$} of 1–2 controlled vortices to guide other vortices \revise{with position $\mathbf{c}_{i}\revise{(t)}$} toward target locations \revise{$\mathbf{c}_i^\text{target}$} via vortex interactions, with the loss defined as $J  = \sum_i\revise{\delta(t-r)}\|\mathbf{c}_i\revise{(t)}-\mathbf{c}_i^\text{target}\|_2^2$, where $\mathbf{c}_i$ denotes the vortex positions and $\mathbf{c}_i^\text{target}$ the target positions. \revise{As $\mathbf{c}_i$ is driven by $\mathbf{u}_t$, which depends on vortex interactions and initial positions, the functional $J$ can be viewed as a functional of $\mathbf{u}_t$ as in \autoref{eq:object_function_def} and optimized over $\mathbf{c}_{j,\text{init}}^{\text{ctrl}}$.} Since controlling vortex relies on preserving vortex structures, flow maps are particularly well-suited for this task. We present the scenarios of \textbf{Single Vortex} case (Fig.~\ref{fig:vortex_control_ex12}), \textbf{Multi-Vortex Single-Target} case and the \textbf{Multi-Vortex Multi-Target} case (Fig.~\ref{fig:vortex_control_ex34}). These examples, with increasing levels of difficulty, demonstrate our method’s ability to preserve vortex structures and to compute accurate gradients for optimization in complex scenarios.
 \textbf{(3) 2D Smoke Control.} This task optimizes a time-varying control force $\mathbf{f}$ to influence the fluid and deform smoke into a target shape \cite{treuille2003keyframe,chen2024fluid}. The force is modeled using $\revise{m}$ ($\revise{m}\sim 3000$) Gaussian wind fields $\mathbf{f}_{i} = \mathbf{w}_ie^{-a\|\mathbf{x}-\mathbf{c}_i\|^2}$ with centers $\mathbf{c}_i$ and strength vector $\mathbf{w}_i$ as parameters, $i=1,..,\revise{m}$ \revise{\cite{treuille2003keyframe}}. The objective function \revise{integrand is $J = \sum_{k\in K}\delta(t - t_k)|\xi(\mathbf{x},t)-\xi^\text{target}(\mathbf{x},t)|^2$}, \revise{where $t_k$ denotes the the time of $k-$th key frame}.
 We showcase continuous keyframe optimization through the \textbf{2D GRAPH Formation} example (Fig.~\ref{fig:2DGRAPH}), the \textbf{2D Life Evolution} example (Fig.~\ref{fig:2Dlife}), and the \textbf{2D Bat Through Obstacles} example (Fig.~\ref{fig:2Dbat}), which involves optimization in the presence of obstacles.  Accurate shape control and fluid realism are enabled by the flow map’s precise passive field mapping and strong advection preservation.  
 \textbf{(4) 3D Smoke Control.} We extend the above task to a 3D volume setting.  We showcase the comparison of \textbf{3D Sphere to Armadillo} (Fig.~\ref{fig:3DArmadillo}) across different resolutions, the continuous 3D shape transitions in \textbf{3D GRAPH} (Fig.~\ref{fig:3DGRAPH}), and the complex topological changes in \textbf{3D Topological Morphing} (Fig.~\ref{fig:3DTopo}). These examples demonstrate that our method remains accurate in 3D, enabling the preservation of fine smoke strands and coherent flow structures.

\begin{table}[t]
    \centering
    \caption{\revise{Task categories and comparison of differentiable methods: differentiable semi-Lagrangian (SL), differentiable Eigenfluids (Eigen), and ours. In the table, $\checkmark$ and $\times$ indicate whether a solver can accomplish the task; bold entries denote the experiments selected for comparison, and underlined entries mark the best-performing solver for each task.}}
    \vspace{-0.25cm}
    \label{tab:experiment_conparison}
    \resizebox{0.99\linewidth}{!}{
    \begin{tabular}{cccccc}
        \toprule
        \textbf{\revise{Task}} & \textbf{\revise{Experiments}} & \textbf{\revise{SL}} & \textbf{\revise{Eigen}} & \textbf{\revise{Ours}} \\
        \midrule
        \revise{Vortex Dynamics Inference} & \revise{Fig.~\textbf{\ref{fig:2DShapeComparisons} (col. 4)}, \ref{fig:2DLogo_and_Gradient}, \ref{fig:2Dobstacal}} & \revise{$\times$} & \revise{$\times$} &  \revise{$\underline{\checkmark}$} \\ 
        \revise{Vortex Control} & \revise{Fig.~\ref{fig:vortex_control_ex12}, \ref{fig:vortex_control_ex34}} & \revise{$\times$}  & \revise{$\times$}  & \revise{$\underline{\checkmark}$}  \\ 
        \revise{Smoke Control} & \revise{Fig.~\textbf{\ref{fig:2DShapeComparisons} (col. 1-3)}, \ref{fig:2Dbat}, \ref{fig:2DGRAPH}, \ref{fig:2Dlife}, \ref{fig:3DTopo}, \ref{fig:3DGRAPH}, \ref{fig:3DArmadillo}} & \revise{$\checkmark$}  & \revise{$\checkmark$}  & \revise{$\underline{\checkmark}$} \\ 
        \bottomrule
    \end{tabular}
    }    
    \vspace{-0.2cm}
\end{table}

\paragraph{Comparison}
We demonstrate the effectiveness of \revise{our adjoint method} in \revise{four experiments} and compare it with baseline methods, including differentiable semi-Lagrangian method \cite{treuille2003keyframe,li2024neuralfluid} and \revise{differentiable} Eigenfluids \cite{chen2024fluid}. For non‑open‑source reasons, \cite{chen2024fluid} is reproduced by adapting the open-source code \cite{borcsok2023control} implementated in $\Phi_{\text{Flow}}$\cite{holl2024phiflow}. 
 \cite{treuille2003keyframe} is implementation by DiffTaichi \cite{hu2019taichi}.  As shown in Fig.~\ref{fig:2DShapeComparisons}, our method demonstrates significantly higher accuracy in both the \textbf{Vortex Dynamics Inference from Images} and \textbf{2D Smoke Control} tasks. \revise{In the 2D smoke control comparison, we further compare with a state-of-the-art method based on the semi-Lagrangian scheme with optimized control forces \cite{tang2021honey}, and our approach still achieves superior results.}  Since neither Eigen Fluid nor Semi-Lagrangian methods preserve vortices, they are unable to accomplish the \textbf{2D \revise{Vortex} Control} task. Moreover, \revise{as shown in \cite{chen2024fluid}}, Eigen Fluid also encounters difficulties in the \textbf{3D Smoke Control} scenario.  \revise{A full summary of task categories and solver comparisons is provided in \autoref{tab:experiment_conparison}.}

\begin{table}[t]
    \centering
    \caption{\revise{Final volume conservation errors (\%) for the 2D smoke control tasks in Fig.~\ref{fig:2DShapeComparisons}. Lower is better.}}
    \vspace{-0.25cm}
    \label{tab:volume_conservation}
    \resizebox{0.99\linewidth}{!}{
    \begin{tabular}{ccccc}
        \toprule
        \textbf{\revise{Volume Percentage}} & \textbf{\revise{Ours}} & \textbf{\revise{DiffEigen}} & \textbf{\revise{Improved-SL}} & 
        \textbf{\revise{SL}}\\
        \midrule
        \revise{Dragon} & \revise{$0.0016\%$} & \revise{$2.89\%$} & \revise{$6.54\%$} &\revise{$0.44\%$}\\ 
        \revise{Snake} & \revise{$0.0056\%$} & \revise{$1.50\%$}  & \revise{$4.91\%$}&\revise{$3.28\%$} \\ 
        \revise{Turtle} & \revise{$0.0018\%$} & \revise{$2.84\%$}  & \revise{$3.46\%$} &\revise{$2.02\%$} \\ 
        \bottomrule
    \end{tabular}
    }    
    \vspace{-0.2cm}
\end{table}
\revise{
\paragraph{Volume Conservation}Benefiting from the accuracy of the flow map formulation, our method preserves smoke volume effectively throughout the control process. As shown in Table~\ref{tab:volume_conservation}, our method achieves fluctuations below $0.006\%$ across all 2D examples in Fig.~\ref{fig:2DShapeComparisons}, whereas competing approaches exhibit deviations ranging from $0.44\%$ to $6.54\%$. In the 3D case (Fig.~\ref{fig:3DGRAPH}), our method further achieves a final-frame fluctuation of only $0.018\%$.}

\paragraph{Time and Memory Cost} We report the runtime and GPU memory usage of our 3D examples to support our choice of numerically solving the adjoint Navier-Stokes equation instead of differentiating the forward process. As shown in \autoref{tab:runtime_stats}, the backward pass has similar runtime and memory cost to the forward pass across different tasks, avoiding the significant increase in backpropagation \revise{that is often attributed} to differentiating complex computation graph.

\begin{table}[t]
    \revise{
    \centering
        \caption{\revise{Runtime (s) and GPU memory cost (GB) comparison between forward and backward passes for different examples. Here, (F) and (B) indicate whether the data corresponds to the forward pass or the backward pass, respectively. The reported times are per step and exclude the computation of external inputs and outputs, such as control forces. Poisson and Advection report the average cost of a single step of solving the Poisson equation and advecting the flow map, respectively. In 2D, we implemented a custom Poisson solver in Taichi, while in 3D we employed the high-performance MGPCG Poisson solver from \cite{Sun_2025}, which makes the 3D Poisson solve significantly faster than its 2D counterpart.}}
    \label{tab:runtime_stats}
    \vspace{-0.25cm}
    \resizebox{0.99\linewidth}{!}{
    \begin{tabular}{cccccccc}
        \toprule
        \textbf{Figure} & \textbf{Resolution} & \textbf{memory (F)} & \textbf{memory (F\&B)} & \textbf{time (F)} & \textbf{time (B)} & \textbf{Poisson} & \textbf{Advection}\\
        \midrule
        Fig.~\ref{fig:2DShapeComparisons} (col. 4) & 256 $\times$ 256 & 1.6 GB & 1.6 GB & 0.17 s & 0.18 s & 0.16 s & 0.004 s\\
        Fig.~\ref{fig:2DLogo_and_Gradient}  & 256 $\times$ 256 & 1.7 GB & 1.7 GB & 0.19 s & 0.20 s & 0.19 s & 0.002 s\\
        Fig.~\ref{fig:2Dobstacal}  & 256 $\times$ 256 & 1.7 GB & 1.7 GB & 0.18 s & 0.21 s & 0.19 s & 0.01 s\\
        Fig.~\ref{fig:2DShapeComparisons} (col. 1-3)  & 256 $\times$ 256 & 1.6 GB & 1.6 GB & 0.16 s & 0.21 s & 0.18 s & 0.01 s\\
        Fig.~\ref{fig:2Dobstacal}  & 256 $\times$ 256 & 1.6 GB & 1.6 GB & 0.17 s & 0.21 s & 0.18 s & 0.01 s\\
        Fig.~\ref{fig:2DGRAPH} \& Fig.~\ref{fig:2Dlife} & 256 $\times$ 256 & 1.6 GB & 1.6 GB & 0.19 s & 0.22 s & 0.20 s & 0.01 s\\
        Fig.~\ref{fig:3DArmadillo} Top  & 196 $\times$ 196 $\times$ 196 & 5.69 GB & 6.53 GB & 0.24 s & 0.28 s & 0.03 s& 0.26 s  \\ 
        Fig.~\ref{fig:3DArmadillo} bottom & 128 $\times$ 128 $\times$ 128 & 2.10 GB  & 2.35 GB  & 0.08 s & 0.09 s & 0.06 s & 0.02 s\\ 
        Fig.~\ref{fig:3DTopo} & 128 $\times$ 128 $\times$ 128 & 2.10 GB  & 2.35 GB  & 0.08 s & 0.08 s & 0.06 s &  0.02 s\\    
        Fig.~\ref{fig:3DGRAPH} & 128 $\times$ 128 $\times$ 128 & 2.10 GB  & 2.35 GB  & 0.09 s & 0.10 s & 0.03 s & 0.05 s \\  
        \bottomrule
    \end{tabular}
    }
    }
\end{table}

\section{Conclusion and Future Work}

This paper presents a differentiable flow map method to improve the accuracy and applicability of differentiable fluid simulation. 
Some limitations still remain. We focus on control forces and velocity fields, without tackling shape optimization and solid boundaries. The differentiation of incompressible flow with a free surface remains to be explored. Future work may also explore shape-based \revise{design} tasks (e.g., \cite{li2024neuralfluid}), \revise{perform shape optimization with real smoke images as targets,} and experiment with more advanced optimization algorithms beyond quasi-Newton methods. 


\begin{acks}
We express our gratitude to the anonymous reviewers for their insightful feedback. Georgia Tech authors acknowledge NSF IIS \#2433322, ECCS \#2318814, CAREER \#2433307, IIS \#2106733, OISE \#2433313, and CNS \#1919647 for funding support.  We credit the Houdini education license for video animations.
\end{acks}




\bibliographystyle{ACM-Reference-Format}
\bibliography{refs_ML_sim.bib, refs_INR.bib, refs_flow_map.bib, refs_simulation.bib, refs_laden.bib,bo.bib, refs_clebsch.bib, refs_adjoint_method.bib, recent_paper.bib}

\clearpage
\appendix


\section*{Supplementary material}

# Supplementary Material: An Adjoint Method for Differentiable Fluid Simulation on Flow Maps

ZHIQI LI*, Georgia Institute of Technology, USA
JINJIN HE*, Georgia Institute of Technology, USA
BARNABÁS BÖRCSÖK, Georgia Institute of Technology, USA
TAIYUAN ZHANG, Dartmouth College, USA
DUOWEN CHEN, Georgia Institute of Technology, USA
TAO DU, Independent Researcher, USA
MING C. LIN, University of Maryland, USA
GREG TURK, Georgia Institute of Technology, USA
BO ZHU, Georgia Institute of Technology, USA

CCS Concepts: • **Computing methodologies → Physical simulation**.

## A Forward Pass Calculation

For the forward pass, we accurately compute Equation 1 using the flow map by reformulating it in a path-integral form, following previous flow map methods [Chen et al. 2025, 2024; Li et al. 2024a,b, 2025a,b; Wang et al. 2025a,b]:

$$
\mathbf{u}(\mathbf{x}, t) = \mathcal{T}_{t \to s}^\top(\mathbf{x})\mathbf{u}(\mathbf{\Psi}_{t \to s}(\mathbf{x}), s) + \mathcal{T}_{t \to s}^\top(\mathbf{x})\mathbf{\Gamma}_{s \to t}(\mathbf{\Psi}_{t \to s}(\mathbf{x})),
$$
$$
\mathbf{\Gamma}_{s \to t}(\mathbf{x}) = \int_s^t \mathcal{F}_{s \to \tau}^\top(\mathbf{x}) \left( -\frac{1}{\rho}\nabla p + \frac{1}{2}\nabla\|\mathbf{u}\|^2 + \mathbf{f} \right)(\mathbf{\Phi}_{s \to \tau}(\mathbf{x}), \tau)d\tau. \tag{20}
$$

The first term $\mathbf{u}_{s \to t}^M(\mathbf{x}) = \mathcal{T}_{t \to s}(\mathbf{x})^\top\mathbf{u}(\mathbf{\Psi}_{t \to s}(\mathbf{x}), s)$ in $\mathbf{u}(\mathbf{x}, t)$ is referred to as the long-range mapped velocity, as it is directly obtained by mapping the accurate initial velocity $\mathbf{u}_s$ through the long-term flow map $\mathbf{\Psi}_{t \to s}$, which prevents the advection of velocity from being affected by accumulated errors. The second term $\mathbf{\Gamma}_{s \to t}$ is called the path integrator, since it accumulates the integral along the trajectory $\mathbf{S}_{\mathbf{x}_0}(t) = \mathbf{\Phi}_{s \to t}(\mathbf{x}_0)$ of any material point $\mathbf{x}_0 \in \mathbb{U}_s$ under the

---

*Joint first author

Authors' Contact Information: Zhiqi Li, zli3167@gatech.edu, Georgia Institute of Technology, USA; Jinjin He, jhe433@gatech.edu, Georgia Institute of Technology, USA; Barnabás Börcsök, borcsok@gatech.edu, Georgia Institute of Technology, USA; Taiyuan Zhang, taiyuan.zhang.gr@dartmouth.edu, Dartmouth College, USA; Duowen Chen, dchen322@gatech.edu, Georgia Institute of Technology, USA; Tao Du, taodu.eecs@gmail.com, Independent Researcher,; Ming C. Lin, lin@umd.edu, University of Maryland, USA; Greg Turk, turk@cc.gatech.edu, Georgia Institute of Technology, USA; Bo Zhu, bo.zhu@gatech.edu, Georgia Institute of Technology, USA.

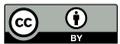
---

flow map. The passive field can also be obtained through long-range mapping based on the flow map

$$
\xi(\mathbf{x}, t) = \xi(\mathbf{\Psi}_{t \to s}(\mathbf{x}), s). \tag{21}
$$

The above computation is generally carried out using the long-short term conversion technique proposed in [Li et al. 2024b]. With $t_c$ denoting the current time and $s$ the initial time:

(1) (**Mapping**) Calculate mapped velocity $\mathbf{u}_{s \to t_c}^M$ as

$$
\mathbf{u}_{s \to t_c}^M = \mathcal{T}_{t_c \to s}^\top(\mathbf{x})\mathbf{u}(\mathbf{\Psi}_{t_c \to s}(\mathbf{x}), s). \tag{22}
$$

(2) (**Conversion**) Convert $\mathbf{u}_{s \to t_c}^M$ to advected velocity $\mathbf{u}_{t_c' \to t_c}^A(\mathbf{x})$ using

$$
\mathbf{u}_{t_c' \to t_c}^A(\mathbf{x}) = \mathbf{u}_{s \to t_c}^M(\mathbf{x}) + \mathcal{T}_{t_c \to s}^\top(\mathbf{x})\mathbf{\Gamma}_{t_c \to s}(\mathbf{\Psi}_{t_c \to s}(\mathbf{x}))
$$
$$
+ \frac{1}{2}\nabla\|\mathbf{u}\|^2\Delta t, \tag{23}
$$

where $t_c'$ is the last step of $t_c$.

(3) (**Force Effect**) Compute the viscous force $\nu\Delta\mathbf{u}$ and external $\mathbf{f}$, and obtain the unprojected velocity as

$$
\mathbf{u}_{t_c}^{up} = \mathbf{u}_{t_c' \to t_c}^A + (\nu\Delta\mathbf{u} + \mathbf{f})\Delta t. \tag{24}
$$

(4) (**Projection**) Perform velocity projection

$$
\mathbf{u}_{t_c} = \mathbf{u}_{t_c}^{up} - \frac{\Delta t}{\rho}\nabla p, \tag{25}
$$

where $p$ is solved from Poisson equation $\frac{\Delta t}{\rho}\Delta p = \nabla \cdot \mathbf{u}_{t_c}^{up}$ with non-through boundary $\mathbf{u}_{t_c} \cdot \mathbf{n} = 0$.

(5) (**Path Integrator Update**) Update the path integrator as

$$
\mathbf{\Gamma}_{s \to t_c}(\mathbf{x}) = \mathbf{\Gamma}_{s \to t_c'}(\mathbf{x}) + \Delta t\mathcal{F}_{s \to t_c}^\top(\mathbf{x})\left(\frac{1}{2}\nabla\|\mathbf{u}\|^2\right.
$$
$$
\left. - \frac{1}{\rho}\nabla p + \nu\Delta\mathbf{u}\right)(\mathbf{\Phi}_{s \to t_c}(\mathbf{x}), t_c). \tag{26}
$$

Similar to the backward pass, the forward pass also suffers from the issue that repeated evolution of $\mathbf{\Psi}$ leads to an $O(m^2)$ time complexity. Therefore, we also apply our long-short time-sparse EFM in the forward pass.

We reinitialize the long-range flow map every $\Delta t_{reinit}^l = n^l\Delta t$ and the short-range flow map every $\Delta t_{reinit}^s = n^s\Delta t$, typically with $n^l = 15 \sim 60$ and $n^s = 1 \sim 3$. In Algorithm 3, we illustrate one long-range reinit cycle starting from the previous long-range reinit time

**Algorithm 3** Long-Short Time-Sparse EFM of Forward Pass

**Initialize:** short-range reinit time $s''$ to long-range reinit time $s'$.

1: **for** each time step $t_c$ between $s'$ and $s' + \Delta t_{\text{reinit}}^l$ **do**
2:    March $\Phi_{s' \to t_c}^l, \Phi_{s'' \to t_c}^s, \mathcal{F}_{s' \to t_c}^l, \mathcal{F}_{s'' \to t_c}^s$ one step;    ▷ eq. 5
3:    **if** $\exists m \in \mathbb{Z}$, st. $t_c = s' + m\Delta t_{\text{reinit}}^s$ **then**
4:      Calculate $\Psi_{t_c \to s''}^s$ and $\mathcal{T}_{t_c \to s''}^s$ by integrating $\Psi_{t_c \to t}^s$ and
5:      $\mathcal{T}_{t_c \to t}^s$ from $t_c$ to $s''$;    ▷ eq. 7
6:      Do mapping and conversion with $\Psi_{t_c \to s''}^s$ and
7:      $\mathcal{T}_{t_c \to s''}^s$;    ▷ eq. 22, 23
8:      Reset $\Phi_{s'' \to t_c}^s$ and $\mathcal{F}_{s'' \to t_c}^s$ and set $s'' = t_c$ ;
9:    **else**
10:      Calculate advection with semi-Lagrangian method;
11:    Calculate force term and projection;
12:    Update $\Gamma_{s' \to t_c}^l$ and $\Gamma_{s'' \to t_c}^s$;    ▷ eq. 26
13:    **if** $t_c = s' + \Delta t_{\text{reinit}}^l$ **then**
14:      Calculate $\Psi_{t_c \to s'}^l$ and $\mathcal{T}_{t_c \to s'}^l$ by integrating $\Psi_{t_c \to t}^l$ and
15:      $\mathcal{T}_{t_c \to t}^l$ from $t_c$ to $s'$;    ▷ eq. 7
16:      Correct results by adding long-range mapping with
17:      $\Psi_{t_c \to s'}^l$, $\mathcal{T}_{t_c \to s'}^l$ and path integrators.    ▷ eq. 20

$s'$. Let $s''$ denote the most recent short-range reinit time, initialized as $s'' = s'$. We maintain two sets of flow maps, $\Phi_{s' \to t}^l, \mathcal{F}_{s' \to t}^l, \Psi_{t_c \to t}^l$, $\mathcal{T}_{t_c \to t}^l$ for long-range, $\Phi_{s'' \to t}^s, \mathcal{F}_{s'' \to t}^s, \Psi_{t_c \to t}^s, \mathcal{T}_{t_c \to t}^s$ for short-range and two path integrators $\Gamma_{s' \to t}^l$ and $\Gamma_{s'' \to t}^s$ for long-range and short-range respectively, where $t_c > s'$ is the current time and $t$ serves as the evolving time variable during integration.

## B Missing Proofs

### B.1 Proof of Equation 9

Similar to [Li et al. 2024b; Nabizadeh et al. 2022], we reformulate Equation 4 as $\left(\frac{\partial}{\partial t} + (\mathbf{u} \cdot \nabla)\right)\mathbf{u}^* + \nabla \mathbf{u}^\top \mathbf{u}^* = 2\nabla \mathbf{u}^\top \mathbf{u}^* + \xi^* \nabla \xi - \frac{1}{\rho} \nabla p^* + \nu \Delta \mathbf{u}^* - \frac{\partial J}{\partial \mathbf{u}}$ and express it in covector form using the Lie derivative

$$\left(\frac{\partial}{\partial t} + \mathcal{L}_{\mathbf{u}}\right)\mathbf{u}^{*\flat} = \lambda^\flat, \tag{27}$$

where $\lambda(\mathbf{x}, t) = \left(2\nabla \mathbf{u}^\top \mathbf{u}^* + \xi^* \nabla \xi - \frac{1}{\rho} \nabla p^* + \nu \Delta \mathbf{u}^* - \frac{\partial J}{\partial \mathbf{u}}\right)(\mathbf{x}, t)$, $\mathcal{L}_{\mathbf{u}} \mathbf{u}^{*\flat}$ denotes the Lie derivative of $\mathbf{u}^{*\flat}$, and $\flat$ is the musical isomorphism mapping a vector to a covector.

By integrating this equation of Lie derivative from $r$ to $t$, we obtain

$$\begin{aligned}
\mathbf{u}_t^{*\flat} &= \Phi_{t \to r}^* \mathbf{u}_r^{*\flat} + \int_r^t (\Psi_{r \to \tau} \circ \Phi_{t \to r})^* \lambda_\tau^\flat d\tau \\
&= \Phi_{t \to r}^* \mathbf{u}_r^{*\flat} + \Phi_{t \to r}^* \int_r^t \Psi_{r \to \tau}^* \lambda_\tau^\flat d\tau,
\end{aligned} \tag{28}$$

where $\Psi_{r \to \tau}^*$ and $\Phi_{t \to r}^*$ are the pullbacks of the covector induced by $\Psi_{r \to \tau}$ and $\Phi_{t \to r}$, respectively. Convert the above expression back to vector form, and note that $\Phi_{r \to \tau}^* \mathbf{v}^\flat$ and $\Psi_{r \to \tau}^* \mathbf{v}^\flat$ corresponds to $\nabla \Phi_{t \to r}^\top \mathbf{v}$ and $\nabla \Psi_{r \to \tau}^\top \mathbf{v}$ respectively for arbitrary vector field $\mathbf{v}$. Then

we get

$$\mathbf{u}^*(\mathbf{x}, t) = \mathcal{F}_{t \to r}^\top(\mathbf{x})\mathbf{u}^*(\Phi_{t \to r}(\mathbf{x}), r) + \mathcal{F}_{t \to r}^\top(\mathbf{x})\Lambda_{r \to t}^u(\Phi_{t \to r}(\mathbf{x})),$$

$$\Lambda_{r \to t}^u(\mathbf{x}) = \int_r^t \mathcal{T}_{r \to \tau}^\top(\mathbf{x})\left(2\nabla \mathbf{u}^\top \mathbf{u}^* + \xi^* \nabla \xi - \frac{1}{\rho} \nabla p + \nu \Delta \mathbf{u}^* \right. \tag{29}$$
$$\left. - \frac{\partial J}{\partial \mathbf{u}}\right)(\Psi_{r \to \tau}(\mathbf{x}), \tau) d\tau,$$

For $\xi^*$, by directly integrating both sides of $\left(\frac{\partial}{\partial t} + (\mathbf{u} \cdot \nabla)\right)\xi^* = -\frac{\partial J}{\partial \xi}$ from $r$ to $t$ along the trajectory $\mathbf{S}_{\mathbf{x}}(\tau) = \Psi_{r \to \tau} \circ \Phi_{t \to r}(\mathbf{x})$, we obtain

$$\xi^*(\mathbf{x}, t) - \xi^*(\Phi_{t \to r}(\mathbf{x}), r) = -\int_r^t \frac{\partial J}{\partial \xi}(\Psi_{r \to \tau} \circ \Phi_{t \to r}(\mathbf{x}), \tau)d\tau \tag{30}$$

which gives the integral representation of $\xi^*$.

### B.2 Proof of Equation 10

Following the approach of [Li et al. 2024b], we prove Equation 10, which establishes the relation between the short-range advected adjoint velocity $\mathbf{u}_{r \to t}^A$ and the long-range mapped adjoint velocity $\mathbf{u}_{r \to t}^{*M}$. First, we decompose $\mathcal{F}_{t \to r}^\top(\mathbf{x})\Lambda_{r \to t}^u(\Phi_{t \to r}(\mathbf{x}))$ as

$$\begin{aligned}
\mathcal{F}_{t \to r}^\top(\mathbf{x})\Lambda_{r \to t}^u(\Phi_{t \to r}(\mathbf{x})) &= \mathcal{F}_{t \to r}^\top(\mathbf{x})\Lambda_{r \to t'}^u(\Phi_{t \to r}(\mathbf{x})) \\
&+ \mathcal{F}_{t \to r}^\top(\mathbf{x})\int_{t'}^t \mathcal{T}_{r \to \tau}^\top(\mathbf{x})\lambda(\Psi_{r \to \tau}(\Phi_{t \to r}(\mathbf{x})), \tau)d\tau,
\end{aligned} \tag{31}$$

where $\lambda(\mathbf{x}, t) = \left(2\nabla \mathbf{u}^\top \mathbf{u}^* + \xi^* \nabla \xi - \frac{1}{\rho} \nabla p^* + \nu \Delta \mathbf{u}^* - \frac{\partial J}{\partial \mathbf{u}}\right)(\mathbf{x}, t)$. Here, the second part $\mathcal{F}_{t \to r}^\top(\mathbf{x})\int_{t'}^t \mathcal{T}_{r \to \tau}^\top(\mathbf{x})\lambda(\Psi_{r \to \tau}(\Phi_{t \to r}(\mathbf{x})), \tau)d\tau$ on the right-hand side of the above equation is equal to $\lambda(\mathbf{x}, t)\Delta t$ when using the Euler scheme to compute with identities $\mathcal{F}_{t \to t'}^\top(\mathbf{x}) = \mathbf{I}$ and $\Psi_{t \to r}(\Phi_{t \to r}(\mathbf{x})) = x$, where $\mathbf{I}$ is the identity matrix. Utilizing the relationship $\mathbf{u}^*(\mathbf{x}, t) = \mathbf{u}_{r \to t}^{*M}(\mathbf{x}) + \mathcal{F}_{t \to r}^\top(\mathbf{x})\Lambda_{r \to t}^u(\Phi_{t \to r}(\mathbf{x})) = \mathbf{u}_{r \to t}^{*A}(\mathbf{x}) + \xi^* \nabla \xi - \frac{1}{\rho} \nabla p^* + \nu \Delta \mathbf{u}^* - \frac{\partial J}{\partial \mathbf{u}}$, we can derive the formula for converting $\mathbf{u}_{r \to t}^{*M}$ to $\mathbf{u}_{t' \to t}^{*A}$ as

$$\mathbf{u}_{t' \to t}^{*A}(\mathbf{x}) = \mathbf{u}_{r \to t}^{*M}(\mathbf{x}) + \mathcal{F}_{t \to r}^\top(\mathbf{x})\Lambda_{r \to t'}^u(\Phi_{t \to r}(\mathbf{x})) + 2\nabla \mathbf{u}^\top \mathbf{u}^* \Delta t. \tag{32}$$

## C Numerical Algorithm for Forward Pass

To provide a clearer illustration of the forward pass, we present its detailed numerical algorithm in Algorithm 4. The discretization is the same as in the backward pass. The computation procedure is generally consistent with [Sun et al. 2025], mainly differing in the use of the Long-Short Time-Sparse EFM method.

In Algorithm 4, the midpoint computation strategy follows [Deng et al. 2023; Sun et al. 2025] and is consistent with that in the backward pass. The viscous term is computed in the same manner as the adjoint viscous term (Equation 18). The term $\left[\nabla \frac{1}{2}\|\mathbf{u}\|_2^2\right]_g$ is evaluated using a similar formulation as in Equation 19, but with a second-order kernel.

---

**Algorithm 4** Long-Short Time-Sparse EFM for Original Field

---

**Initialize:** $\mathbf{u}_{s',g}$, $\mathbf{u}_{s'',g}$ to initial velocity; $\mathcal{T}_{t_c \to s'',g}^s$, $\mathcal{F}_{s'' \to t_c,g}^s$, $\mathcal{T}_{t_c \to s',g}^l$, $\mathcal{F}_{s' \to t_c,g}^l$ to $\mathbf{I}$; $\Psi_{t_c \to s'',g}^s$, $\Phi_{s'' \to t_c,g}^s$, $\Psi_{t_c \to s',g}^l$, $\Phi_{s' \to t_c,g}^l$ to $\mathbf{x}_g$; $\Gamma_{s' \to t_c,g}^l$ and $\Gamma_{s'' \to t_c,g}^s$ to 0; $s'$, $s''$, $t_c$ to $s$.

---

1: **for** each time step $t_c$ from $t_0$ to $t_n$ **do**
2:      Calculate midpoint velocity $\mathbf{u}_g^{mid}$;
3:      March $\Phi_{s' \to t_c,g}^l, \Phi_{s'' \to t_c,g}^s, \mathcal{F}_{s' \to t_c,g}^l, \mathcal{F}_{s'' \to t_c,g}^s$ one step; ▷ eq. 5
4:      **if** $c \pmod{n^s} = 0$ **then**
5:          Integrate $\mathcal{T}_{t_c \to s'',g}^s$ and $\Psi_{t_c \to s'',g}^s$ from $t_c$ to $s''$; ▷ eq. 7
6:          Calculate mapped velocity $\mathbf{u}_{s'' \to t_c,g}^M$ and convert to one-step advected velocity $\mathbf{u}_{t_{c-1} \to t_c,g}^A$; ▷ eq. 22, 23
7:          Set initial time for short mapping $s''$ to $t_c$;
8:          Reinitialize $\mathcal{F}_{s'' \to t_c,g}^s$ to $\mathbf{I}$, $\Phi_{s'' \to t_c,g}^s$ to $\mathbf{x}_g$;
9:          Calculate $\xi_{t_c \to s'',g}$ by $\Psi_{t_c \to s'',g}^s$; ▷ eq. 21
10:     **else**
11:          Calculate $\mathbf{u}_{t_{c-1} \to t_c,g}^A$ and $\xi_{t_{c-1} \to t_c,g}^A$ by semi-Lagrangian;
12:      Compute viscous term $[\nu \Delta \mathbf{u}]_g$, force term $\mathbf{f}_g$ and $[\nabla \frac{1}{2} \|\mathbf{u}\|_2^2]_g$ on the grid;
13:      Compute unprojected velocity $\mathbf{u}_{t_c,g}^{up}$; ▷ eq. 24
14:      Calculate $\mathbf{u}_{t_c,g}$ using $p$ from Poission equation; ▷ eq. 25
15:      Update both short and long path integrator $\Gamma_{s' \to t_c,g}^l$ and $\Gamma_{s'' \to t_c,g}^s$; ▷ eq. 26
16:      **if** $c \pmod{n^l} = 0$ **then**
17:          Integrate $\mathcal{T}_{t_c \to s',g}^l$ and $\Psi_{t_c \to s',g}^l$ from $t_c$ to $s'$; ▷ eq. 7
18:          Calculate accurate $\mathbf{u}_{t_c,g}$ by mapping with $\mathcal{T}_{t_c \to s',g}^l$, $\Psi_{t_c \to s',g}^l$ and integrator $\Lambda_{r' \to s',g}^{u,l}$; ▷ eq. 20
19:          Set initial time for long mapping $s'$ to $t_c$.

---



\end{document}